\documentclass[onecolumn,useAMS,usenatbib]{mn2e}

\usepackage[dvips,draft]{graphicx}

\title[Radiation hydrodynamics in Kerr spacetime]
{Radiation Hydrodynamics in Kerr Spacetime: Equations without 
Coordinate Singularity at the Event Horizon}
\author[Rohta Takahashi]{Rohta Takahashi$^{1,2}$\thanks{E-mail:
rohta@ea.c.u-tokyo.ac.jp}\\
$^{1}$Graduate School of Arts and Sciences, University of Tokyo, Komaba, 
Meguro, Tokyo 153-8902, Japan
\\
$^{2}$Montana State University, 
Department of Physics, Bozeman, MT 59717-0350
}

\begin{document}

\date{Accepted 200X December 15. Received 200X December 14; 
in original form 200X October 11}

\pagerange{\pageref{firstpage}--\pageref{lastpage}} \pubyear{2007}

\maketitle

\label{firstpage}

\begin{abstract}
Equations of fully general relativistic radiation hydrodynamics 
around a rotating black hole 
are derived by using the Kerr-Schild coordinate 
where there is no coordinate singularity at the event horizon.   
Since the radiation interacts with matter moving with relativistic 
velocities near the event horizon, the interplay between the radiation 
and the matter should be described fully relativistically. 
In the formalism used in this study,  
while the interactions between matter and radiation are introduced 
in the comoving frame, the equations and the equations and the derivatives 
for the description of the global evolution of both matter and the 
radiation are given in the Kerr-Schild frame (KSF) which is a frame 
fixed to the coordinate describing the central black hole. 
As a frame fixed to the coordinate, 
we use the locally non-rotating reference frame (LNRF) 
representing a radially falling frame when the Kerr-Schild coordinate 
is used. 
Around the rotating black hole, both the matter and the radiation 
are affected by the frame-dragging effects.  
\end{abstract}


%%%%%%%%%%%%%%%%%%%%%%%%%%%%%%%%%%%%%%%%%%%%%%%%%%%%%%%%%%%%%%%%%%%%%%%%%%%
%%%%%%%%%%%%%%%%%%%%%%%%%%%%%%%%%%%%%%%%%%%%%%%%%%%%%%%%%%%%%%%%%%%%%%%%%%%
\section{Introduction}
It is widely believed that the accretion flow onto black holes plays one of 
the essential roles in the active phenomena in the universe. 
Relativistic effects in accretion disks around black holes 
derives some major activities of astrophysical black holes, such as 
active galactic nuclei, Galactic black hole candidates and possibly 
gamma-ray bursts. 
For the mass accretion rate near or over the Eddington mass accretion 
rate, the interactions between radiation and matter in the accretion disk 
are important. 
While 
for the supercritical accretion flows considered in e.g. Syfert galaxies 
or the black hole binaries the photons interact with matter,  
for the hypercritical accretion flows considered in 
the central engine of gamma-ray bursts the neutrinos interact 
with matter around the central black hole. 
In such situations, the dynamics and the energy balance in matter 
and radiation are affected by each other. 
So far, the general relativistic radiative transfer is investigated 
by many authors 
(e.g. Lindquist 1966; Anderson \& Spiegel 1972; Schmid-Burgk 1978; 
Thorne 1981; Schinder, Bludman \& Piran 1988; Turolla \& Nobili 1988; 
Anile \& Romano 1992; Cardall \& Mezzacappa 2003; Park 2006).  
\cite{L66} gives a general treatment of the radiation transfer equation and 
the radiation hydrodynamic equations derived by using a comoving Lagrangian 
frame of reference. 
\cite{T81} derive general relativistic moment equations 
up to an arbitrary order by introducing projected symmetric 
trace-free (PSTF) tensors. 
Since these formalisms are based on the comoving frames, 
the physical quantities in terms of matter and radiation and 
the directional derivatives are also described in the comoving frame. 
In some astrophysical objects, the radiation interacts with matters 
moving at relativistic velocities. 
In such cases, the interaction between the radiation and the matter 
should be described fully relativistically. 
This is easily done if both radiation field and matters are evaluated 
in the comoving frame which is a frame where the element of matter is at 
rest. 
Based on this idea, \cite{M80} introduce 
the radiation hydrodynamic equations in the Eulerian framework.    
In this approach, the physical quantities in terms of 
matter and radiation and the derivatives were introduced in the frame 
fixed to the coordinate of the central object, e.g. a black hole, 
while the interactions between matter and radiation were calculated 
in the comoving frame. 
That is,  
the local processes like the interaction between the matter and the 
radiations are evaluated in the comoving frame, while 
the derivatives which are used for the calculations of the global 
dynamics of both the matter and the radiation are derived in the 
frame fixed to the coordinate describing the central object.  
In principle, this formalism can be extended to arbitrary space-time. 
\cite{P93} derived the radiation hydrodynamic equations for 
spherically symmetric systems, and \cite{P06} give the explicit 
expressions for the basic equations of the general relativistic 
radiation hydrodynamics in Schwarzschild space-time.   
In addition, the equations of the general relativistic radiation 
hydrodynamics in Kerr space-time are given by \cite{T07b} based 
on the Boyer-Lindquist coordinate which is one of the traditional 
coordinate of the metric of a rotating black hole and is most frequently 
used in the past studies. 
However, it is known that the Boyer-Lindquist coordinate have 
the coordinate singularity at the event horizon. 
Due to this coordinate singularity, the flow structure calculated 
by using this coordinate exhibit some unrealistic behavior near the 
event horizon.
For example, the accretion flow plunges into the horizon with the speed 
of light, and the angular velocity of the accretion flow become equal to 
the angular velocity of the frame-dragging at the event horizon 
(e.g. Takahashi 2007a). 
Both these features are caused by the coordinate singularity, and 
the actual flows near the horizon do not have such properties. 
So, the alternative coordinate for the description 
of the metric around a rotating black hole are sometimes used 
mainly in the dynamical hydrodynamic or magnetohydrodynamic simulations 
\citep{pf98, fip98, c00, f00, k01, gmt03, k04, gsm04}. 
One of such coordinates is the Kerr-Schild coordinate where 
there is no coordinate singularity at the event horizon. 
By using this coordinate, 
the structure of the accretion flow can be calculated 
accurately near or just on the event horizon. 
So, since the inner boundary condition can be set inside the horizon, 
the equations of the radiation hydrodynamics around a 
rotating black hole based on the Kerr-Schild coordinate become useful 
when the numerical calculations are performed. 
One of the main purposes in this paper is to give these equations. 
After giving the general forms of the basic equations for 
the general relativistic radiation hydrodynamics in \S 2, in \S 3 
by using the orthonormal tetrads fixed to the coordinate (\S 3.1),  
the radiation moments (\S 3.2), the radiation four-force (\S 3.3) 
are derived. 
In \S 4, we also give the basic equations for the radiation 
hydrodynamics including the continuity equation (\S 4.1), the hydrodynamic 
equations (\S 4.2) and the radiation moment equations (\S 4.3). 
%
%Discussion is given in \S 4 and 
Concluding remarks are given in the last section. 
In this paper, we assume $c=1$ in most equations except in a few cases 
where $c$ is explicitly used for clarity. 
Latin and Greek indices denote spatial components and spacetime components, 
respectively. 
$\nabla_\mu$ denote the covariant derivative with respect to $g_{\mu\nu}$. 
%

%%%%%%%%%%%%%%%%%%%%%%%%%%%%%%%%%%%%%%%%%%%%%%%%%%%%%%%%%%%%%%%%%%%%%%%%%%%
%%%%%%%%%%%%%%%%%%%%%%%%%%%%%%%%%%%%%%%%%%%%%%%%%%%%%%%%%%%%%%%%%%%%%%%%%%%
\section{Covariant equations 
for general relativistic radiation hydrodynamics}
Before the explicit expressions for the equations of 
the radiation hydrodynamics in Kerr space-time are given, 
here we briefly summarize the the basic equations of the general 
relativistic radiation hydrodynamics in covariant form 
(e.g. Mihalas \& Miharas 1984). 
We assume the energy-momentum tensor for matter of an ideal gas 
described as 
%
%%%%%
\begin{equation}
T^{\alpha\beta}=\rho_0 h_{\rm g}u^\alpha u^\beta +P_{\rm g}g^{\alpha\beta},
\end{equation}
%%%%%
%
where $u^\alpha$, $\rho_0$, $h_{\rm g}$ and $P_{\rm g}$ are 
the four-velocity, the rest-mass density, 
the relativistic specific enthalpy and the pressure of the gas, 
respectively. 
The specific enthalpy of the gas, $h_{\rm g}$, is calculated as 
$h_{\rm g}=(\varepsilon_{\rm g}+P_{\rm g})/\rho_0$ 
where $\varepsilon_{\rm g}$ is the energy density of the gas. 
Here, the fluid quantities 
$h_{\rm g}$, $\varepsilon_{\rm g}$, $P_{\rm g}$ and $\rho_0$ are 
all being measured in the comoving frame of the fluid. 
The radiation stress-energy tensor is calculated from the 
specific intensity $I(x^\alpha;{\bf n},\nu)$ as 
%
%%%%%
\begin{equation}
R^{\alpha\beta}=\int\int I({\bf n},\nu)n^\alpha n^\beta d\nu d\Omega,  
\end{equation}
%%%%%
%
where $p^\alpha=(h\nu)(1,{\bf n})$ are the photon (or neutrino) 
four-momentum and $n^\alpha\equiv p^\alpha/h\nu$. 
Here, the specific intensity $I(x^\alpha;{\bf n},\nu)$ is defined for 
photons (or neutrinos) moving in direction ${\bf n}$ 
with the frequency $\nu$. 
The particle number conservation equation in the absence of 
particle creation and annihilation and  
the conservation equation for the total energy-momentum 
of gas plus radiation are given as  
%
%%%%%
\begin{eqnarray}
%(nu^\alpha)_{;\alpha}&=&0, \\
%(T^{\alpha\beta}+R^{\alpha\beta})_{;\beta}&=&0,  
\nabla_\alpha(nu^\alpha)&=&0, \\
\nabla_\beta (T^{\alpha\beta}+R^{\alpha\beta})&=&0,  
\end{eqnarray}
%%%%%
%
respectively. 
Here $n$ is the particle number density measured in the comoving frame, 
and $n$ is related to $\rho_0$ as $n=\rho_0/m_{\rm g}$ where 
$m_{\rm g}$ is the mass of the gas particle.  
On the other hand, 
the radiation four-force density acting on the matter is given as 
%
%%%%%
\begin{equation}
G^\alpha=\frac{1}{c}\int \int [\chi_\nu I({\bf n},\nu)-\eta_\nu]n^\alpha 
d\nu d\Omega ,
\end{equation}
%%%%%
%
where $\chi_\nu$ and $\eta_\nu$ are the opacity and the emissivity, 
respectively.  
The invariant emissivity and invariant opacity are $\eta_\nu/\nu^2$ and 
$\nu\chi_\nu$, respectively. 
The dynamical equations for the matter and the radiation field are 
described as 
%
%%%%%
\begin{eqnarray}
%T^{\alpha\beta}_{~~~;\beta}&=& G^\alpha,\\
%R^{\alpha\beta}_{~~~;\beta}&=&-G^\alpha.   
\nabla_\beta T^{\alpha\beta}&=& G^\alpha,\\
\nabla_\beta R^{\alpha\beta}&=&-G^\alpha.   
\end{eqnarray}
%%%%%
%

%%%%%%%%%%%%%%%%%%%%%%%%%%%%%%%%%%%%%%%%%%%%%%%%%%%%%%%%%%%%%%%%%%%%%%%%%%%
%%%%%%%%%%%%%%%%%%%%%%%%%%%%%%%%%%%%%%%%%%%%%%%%%%%%%%%%%%%%%%%%%%%%%%%%%%%
\section{Radiation Hydrodynamics described by Kerr-Schild Coordinate}
In this study, we assume 
the background geometry around the rotating black hole written by 
the Kerr-Schild coordinate where there is no coordinate singularity
at the event horizon. 
By using this coordinate, the metric around a rotating black hole 
is described as 
%
%%%%%
\begin{eqnarray}
ds^2
&=&g_{\alpha\beta}dx^\alpha dx^\beta,\nonumber\\
&=&-\alpha^2 dt^2 +\gamma_{ij}(dx^i+\beta^i dt)(dx^j+\beta^j dt), 
\end{eqnarray}
%%%%%
%
where $i$, $j=r$, $\theta$, $\phi$ and the nonzero components of the lapse 
function $\alpha$, the shift vector $\beta^i$ 
and the metric in three-dimensional spatial hypersurface $\gamma_{ij}$ 
are given in the geometric unit as 
%
%%%%%%
%%%\begin{mathletters}
\begin{eqnarray}
\alpha&=&\left(1+\frac{2mr}{\Sigma}\right)^{-1/2},~
\beta^r=\frac{2mr/\Sigma}{1+2mr/\Sigma},
\nonumber\\
\gamma_{rr}&=&1+\frac{2mr}{\Sigma},
\gamma_{\theta\theta}=\Sigma,~~
\gamma_{\phi\phi}=\frac{A\sin^2\theta}{\Sigma},
\gamma_{r\phi}=\gamma_{\phi r}
=-a\sin^2\theta\left(1+\frac{2mr}{\Sigma}\right).
\end{eqnarray}
%%%\end{mathletters}
%%%%%%
%
Here, we use the geometric mass $m=GM/c^2$, 
$\Sigma=r^2+a^2\cos^2\theta$,  
$A=(r^2+a^2)^2-a^2\Delta\sin^2\theta
=\Sigma\Delta+2mr(r^2+a^2)
=\Sigma^2 + a^2 \sin^2\theta (\Sigma + 2mr)$, and 
$\Delta=r^2-2Mr+a^2$, 
where $M$ is the black hole mass, $G$ is the gravitational constant and 
$c$ is the speed of light. 
The position of the outer and inner horizon, $r_\pm$, 
is calculated from $\Delta=0$ 
as $r_\pm=m\pm(m^2-a^2)^{1/2}$.    
The angular velocity of the frame dragging due to the black hole's 
rotation is calculated as $\omega=-g_{t\phi}/g_{\phi\phi}=2mar/A$. 
%

%%%%%%%%%%%%%%%%%%%%%%%%%%%%%%%%%%%%%%%%%%%%%%%%%%%%%%%%%%%%%%%%%%%%%%%%%%%
\subsection{Reference Frames and Orthonormal Tetrads}
Three reference frames are used in this study: 
(1) the Kerr-Schild frame (KSF) which is the frame 
based on the Kerr-Schild coordinate describing the metric, 
(2) the locally non-rotating reference frame (LNRF) 
which is the orthonormal frame fixed to the coordinate,  
and 
(3) the comoving frame where the element of the fluid is at rest. 
The interactions between the matter 
and the radiation are introduced in the comoving frame.  
%
%The physical quantities measured in the LNRF and the comoving frame 
%are denoted by the hat and the bar, respectively. 
%
The LNRF is calculated from a stationary congruence formed by observers 
with a future-directed unit vector orthogonal to $t=$constant. 
The components of the four velocity for such observer are given as 
\citep{FN98}
%
%%%%%
\begin{equation}
u_\mu=-\alpha\delta^t_\mu,~~~{\rm and}~~~
%\end{equation}
%and 
%\begin{equation}
u^{t}=\alpha^{-1},~~
u^{i}= -\alpha^{-1}\beta^i,~~(i=r,~\theta,~\phi),   
\end{equation}
%%%%%
%
respectively. 
Since the vorticity tensor vanishes for this congruences, 
the reference frame formed by this congruence is locally non-rotating. 
Then, the tetrad vectors, $e^{\hat{\mu}}_{\nu}$ 
and $e^\mu_{\hat{\nu}}$, for the LNRF are given as 
%
%%%%%
\begin{eqnarray}
e^{\hat{t}}_{\mu}      
	&=& \left[\alpha,~0,~0,~0\right],\\
e^{\hat{r}}_{\mu}      
	&=& \left[\beta^r(\gamma^{rr})^{-1/2},~(\gamma^{rr})^{-1/2},
		~0,~0\right],\\
e^{\hat{\theta}}_{\mu} 
	&=& \left[0,~0,~(\gamma_{\theta\theta})^{1/2},~0\right],\\
e^{\hat{\phi}}_{\mu}   
	&=& \left[\beta^r\gamma_{r\phi}(\gamma_{\phi\phi})^{-1/2},
		~\gamma_{r\phi}(\gamma_{\phi\phi})^{-1/2},
		~0,
		~(\gamma_{\phi\phi})^{1/2}
		\right], 
\end{eqnarray}
%%%%%
%
and 
%
%%%%%
\begin{eqnarray}
e^\mu_{\hat{t}}     &=&
	\left[\alpha^{-1},~-\beta^r\alpha^{-1},~0,~0\right],\\
e^\mu_{\hat{r}}     &=&
	\left[0,~(\gamma^{rr})^{1/2},
		~0,~\gamma^{r\phi}(\gamma^{rr})^{-1/2}\right],\\
e^\mu_{\hat{\theta}}&=&
	\left[0,~0,~(\gamma^{\theta\theta})^{1/2},~0\right],\\
e^\mu_{\hat{\phi}}  &=&
	\left[0,~0,~0,~(\gamma_{\phi\phi})^{-1/2}\right]. 
\end{eqnarray}
%%%%%
%
Here, the hat denotes the physical quantities measured in the LNRF. 
The base ${\bf e}_{\hat{\alpha}}=\partial/\partial x^{\hat{\alpha}}$ 
of the LNRF can be expressed by the coordinate base 
$\partial/\partial x^\mu$ as 
%
%%%%%
\begin{equation}
\frac{\partial}{\partial\hat{t}}= 
	\frac{1}{\alpha}\left(\frac{\partial}{\partial t}
	-\beta^r \frac{\partial}{\partial r}\right),~~~
\frac{\partial}{\partial\hat{r}}= 
	\sqrt{\mathstrut \gamma^{rr}}
	\left(
		\frac{\partial}{\partial r}-
		\frac{\gamma_{r\phi}}{\gamma_{\phi\phi}}
			\frac{\partial}{\partial \phi}
	\right),~~~
\frac{\partial}{\partial\hat{\theta}}= 
	\frac{1}{\sqrt{\mathstrut \gamma_{\theta\theta}}}
	\frac{\partial}{\partial \theta},~~~
\frac{\partial}{\partial\hat{\phi}}= 
	\frac{1}{\sqrt{\mathstrut \gamma_{\phi\phi}}}
	\frac{\partial}{\partial \phi}, 
\end{equation}
%%%%%
%
where $\mu=t,~r,~\theta,~\phi$. 
Here, $(x^{\hat{0}},~x^{\hat{1}},~x^{\hat{2}},~x^{\hat{3}})
=(\hat{t},~\hat{r},~\hat{\theta},~\hat{\phi})$. 
The components of the fluid's three velocity measured by 
a fiducial observer who is fixed with respect to the coordinates in the LNRF 
are calculated as 
%
%%%%%
\begin{equation}
\hat{v}^i=\frac{u^{\hat{i}}}{u^{\hat{t}}},~~~(i=r,~\theta,~\phi)
\end{equation}
%%%%%
%
where $u^{\hat{\alpha}}$ is the four-velocity in the LNRF. 
The components of the three velocity are explicitly 
calculated as 
%
%%%%%
\begin{equation}
\hat{v}^r=\frac{1}{\alpha \sqrt{\mathstrut \gamma^{rr}}}
	\left(
		\frac{u^r}{u^t}+\beta^r
	\right),~~~
\hat{v}^\theta
	=\frac{1}{\alpha \sqrt{\mathstrut \gamma^{\theta\theta}}}
	\left(
		\frac{u^\theta}{u^t}
	\right),~~~
\hat{v}^r=\frac{\sqrt{\mathstrut \gamma_{\phi\phi}}}{\alpha}
	\left[\Omega + 
			\frac{\gamma_{r\phi}}{\gamma_{\phi\phi}}
			\left(
				\frac{u^r}{u^t}+\beta^r
			\right)
	\right],~~~
\end{equation}
%%%%%
%
where $\Omega\equiv u^\phi/u^t$ is the angular velocity and 
we have used $u^{\hat{t}}=-u_{\hat{t}}=\alpha u^t$. 
The Lorentz factor $\hat{\gamma}$ for this three velocity is 
calculated as 
%
%%%%%
\begin{equation}
\hat{\gamma}\equiv(1-\hat{v}^2)^{-1/2}=\alpha u^t,  
\end{equation}
%%%%%
%
where $\hat{v}^2={\bf v}\cdot{\bf v}=\hat{v}_i\hat{v}^i=\hat{v}_r^2+\hat{v}_\theta^2+\hat{v}_\phi^2$. 

A tetrad base for the comoving frame 
$\partial/\partial x^{\bar{\alpha}}$
is calculated by 
the Lorentz transformation as 
%
%%%%%
\begin{equation}
\frac{\partial}{\partial x^{\bar{\alpha}}}
=\Lambda^{\hat{\beta}}_{~\bar{\alpha}}({\bf v})
\frac{\partial}{\partial x^{\hat{\beta}}}, 
\end{equation}
%%%%%
%
where the bar denotes the physical quantities measured 
in the comoving frame. 
The components of the Lorentz transformation 
$\Lambda^{\hat{\alpha}}_{~\bar{\beta}}({\bf v})$ are given as 
$\Lambda^{\hat{t}}_{~\bar{t}}=\hat{\gamma}$, 
$\Lambda^{\hat{i}}_{~\bar{t}}=\hat{\gamma} \hat{v}^i$, 
$\Lambda^{\hat{t}}_{~\bar{j}}=\hat{\gamma} \hat{v}_j$ and 
$\Lambda^{\hat{i}}_{~\bar{j}}=\delta^i_{~j}+\hat{v}^i \hat{v}_j \hat{\gamma}^2/(1+\gamma)$ 
($i,~j=r,~\theta,~\phi$).   
Here, $(x^{\bar{0}},~x^{\bar{1}},~x^{\bar{2}},~x^{\bar{3}})
=(\bar{t},~\bar{r},~\bar{\theta},~\bar{\phi})$. 
The components of the base of the comoving tetrad 
$\partial/\partial x^{\bar{\alpha}}$
can be expressed by the coordinate base 
$\partial/\partial x^{\alpha}$
as 
%
%%%%%
\begin{eqnarray}
\frac{\partial}{\partial\bar{t}}&=&
	\frac{\hat{\gamma}}{\alpha} \frac{\partial}{\partial t}
	+\hat{\gamma}
		\left(
			\hat{v}_r \sqrt{\mathstrut \gamma^{rr}}-\frac{\beta^r}{\alpha}
		\right)
		\frac{\partial}{\partial r}
	+\frac{\hat{\gamma} \hat{v}_\theta}
			{\sqrt{\mathstrut \gamma_{\theta\theta}}}
		\frac{\partial}{\partial \theta}
	+\hat{\gamma}\left(
		\frac{ \hat{v}_\phi}
			{\sqrt{\mathstrut \gamma_{\phi\phi}}}
		-\hat{v}_r \sqrt{\mathstrut \gamma^{rr}}
			\frac{\gamma_{r\phi}}{\gamma_{\phi\phi}}
	\right)
		\frac{\partial}{\partial \phi},\nonumber\\
\frac{\partial}{\partial\bar{r}}&=&
	\frac{\hat{\gamma} \hat{v}_r}{\alpha} \frac{\partial}{\partial t}
	+\left[
		\sqrt{\mathstrut \gamma^{rr}}
			\left(
				1+
				\frac{\hat{v}_r^2\hat{\gamma}^2}{\hat{\gamma}+1}
			\right)
		-\hat{\gamma}\hat{v}_r\frac{\beta^r}{\alpha}
	\right] 
		\frac{\partial}{\partial r}
	+\frac{1}{\sqrt{\mathstrut \gamma_{\theta\theta}}}
		\left(
			\frac{\hat{v}_r \hat{v}_\theta\hat{\gamma}^2}{\hat{\gamma}+1}
		\right)
		\frac{\partial}{\partial \theta}
\nonumber\\
&&
	+\left[
		\frac{1}{\sqrt{\mathstrut \gamma_{\phi\phi}}}
		\left(
			\frac{\hat{v}_r \hat{v}_\phi\hat{\gamma}^2}{\hat{\gamma}+1}
		\right)
		-\sqrt{\mathstrut \gamma^{rr}}
		\frac{\gamma_{r\phi}}{\gamma_{\phi\phi}}
		\left(
			1+\frac{\hat{v}_r^2\hat{\gamma}^2}{1+\hat{\gamma}}
		\right)
	\right]
		\frac{\partial}{\partial \phi},\nonumber\\
\frac{\partial}{\partial\bar{\theta}}&=&
	\frac{\hat{\gamma} \hat{v}_\theta}{\alpha} \frac{\partial}{\partial t}
	+\hat{\gamma}\hat{v}_\theta
	\left[
		\sqrt{\mathstrut \gamma^{rr}}
		\left(
			\frac{\hat{v}_r \hat{\gamma}}{\hat{\gamma}+1}
		\right)
		-\frac{\beta^r}{\alpha}
	\right]
	\frac{\partial}{\partial r}
	+\frac{1}{\sqrt{\mathstrut \gamma_{\theta\theta}}}
		\left(
			1+\frac{\hat{v}_\theta^2\hat{\gamma}^2}{\hat{\gamma}+1}
		\right) 
		\frac{\partial}{\partial \theta}
	+\frac{\hat{v}_\theta\hat{\gamma}^2}{\hat{\gamma}+1}
	\left(
		\frac{\hat{v}_\phi}{\sqrt{\mathstrut \gamma_{\phi\phi}}}
		-\hat{v}_r\sqrt{\mathstrut \gamma^{rr}}
			\frac{\gamma_{r\phi}}{\gamma_{\phi\phi}}
	\right)
	\frac{\partial}{\partial \phi},\nonumber\\
\frac{\partial}{\partial\bar{\phi}}&=&
	\frac{\hat{\gamma} \hat{v}_\phi}{\alpha} \frac{\partial}{\partial t}
	+\hat{\gamma}\hat{v}_\phi
		\left[
			\sqrt{\mathstrut \gamma^{rr}}
			\left(
				\frac{\hat{v}_r\hat{\gamma}}{\hat{\gamma}+1}
			\right)
			-\frac{\beta^r}{\alpha}
		\right]
		\frac{\partial}{\partial r}
	+\frac{1}{\sqrt{\mathstrut \gamma_{\theta\theta}}}
		\left(
			\frac{\hat{v}_\theta \hat{v}_\phi\hat{\gamma}^2}{\hat{\gamma}+1}
		\right)
		\frac{\partial}{\partial \theta}
\nonumber\\
&&
	+\left[
		\frac{1}{\sqrt{\mathstrut \gamma_{\phi\phi}}}
		\left(
			1+\frac{\hat{v}_\phi^2\hat{\gamma}^2}{\hat{\gamma}+1}
		\right)
		-\sqrt{\mathstrut \gamma^{rr}}
			\frac{\gamma_{r\phi}}{\gamma_{\phi\phi}}
		\left(
		\frac{\hat{v}_r\hat{v}_\phi\hat{\gamma}^2}{\hat{\gamma}+1}
		\right)
	\right]
		\frac{\partial}{\partial \phi}.
\end{eqnarray}
%%%%%
%
In the similar manner, the inverse transformation from the base of 
the comoving tetrad to the coordinate base is calculated by using 
the inverse Lorentz transformation 
$\Lambda^{\bar{\alpha}}_{~\hat{\beta}}(-{\bf v})$, and 
the coordinate base $\partial/\partial x^{\alpha}$ 
is calculated from the base of the comoving tetrad 
$\partial/\partial x^{\bar{\alpha}}$ as 
% 
%%%%%
\begin{eqnarray}
\frac{1}{\alpha}
\frac{\partial}{\partial t}&=&
\hat{\gamma} 
		\left[
			1-
			\frac{\beta^r}{\alpha}
			\left(
				\frac{\hat{v}_r}{\sqrt{\mathstrut \gamma^{rr}}}
				-\frac{\hat{v}_\phi\gamma_{r\phi}}
					{\sqrt{\mathstrut \gamma_{\phi\phi}}}
			\right)
		\right]
		\frac{\partial}{\partial \bar{t}}
	+\left[
		-\hat{\gamma}\hat{v}_r
		+\frac{\beta^r}{\alpha\sqrt{\mathstrut \gamma^{rr}}}
		\left(
			1+
			\frac{\hat{\gamma}^2\hat{v}_r^2}{\hat{\gamma}+1}
		\right)
		+\frac{\beta^r\gamma_{r\phi}}
			{\alpha\sqrt{\mathstrut\gamma_{\phi\phi}}}
		\left(
			\frac{\hat{\gamma}^2\hat{v}_r \hat{v}_\phi}{\hat{\gamma}+1}
		\right)
	\right]
		\frac{\partial}{\partial \bar{r}}
\nonumber\\
&&
	+\left[
		-\hat{\gamma}\hat{v}_\theta
		+\frac{\beta^r}{\alpha\sqrt{\mathstrut \gamma^{rr}}}
		\left(
			\frac{\hat{\gamma}^2\hat{v}_r\hat{v}_\theta}{\hat{\gamma}+1}
		\right)
		+\frac{\beta^r\gamma_{r\phi}}
			{\alpha\sqrt{\mathstrut\gamma_{\phi\phi}}}
		\left(
			\frac{\hat{\gamma}^2\hat{v}_\theta\hat{v}_\phi}{\hat{\gamma}+1}
		\right)
	\right]
		\frac{\partial}{\partial \bar{\theta}}
\nonumber\\
&&
	+\left[
		-\hat{\gamma}\hat{v}_\phi
		+\frac{\beta^r}{\alpha\sqrt{\mathstrut \gamma^{rr}}}
		\left(
			\frac{\hat{\gamma}^2\hat{v}_r\hat{v}_\phi}{\hat{\gamma}+1}
		\right)
		+\frac{\beta^r\gamma_{r\phi}}
			{\alpha\sqrt{\mathstrut\gamma_{\phi\phi}}}
		\left(
			1+
			\frac{\hat{\gamma}^2\hat{v}_\phi^2}{\hat{\gamma}+1}
		\right)
	\right]
		\frac{\partial}{\partial \bar{\phi}},\nonumber\\
\sqrt{\mathstrut \gamma^{rr}}
\frac{\partial}{\partial r}&=&
	-\hat{\gamma} 
		\left(
			\hat{v}_r
			+\hat{v}_\phi\gamma_{r\phi}
			\sqrt{\frac{\gamma^{rr}}{\gamma_{\phi\phi}}}
		\right)
		\frac{\partial}{\partial \bar{t}}
	+\left[
		\left(
			1+
			\frac{\hat{\gamma}^2\hat{v}_r^2}{\hat{\gamma}+1}
		\right)
		+\gamma_{r\phi}\sqrt{\frac{\gamma^{rr}}{\gamma_{\phi\phi}}}
		\left(
			\frac{\hat{\gamma}^2\hat{v}_r \hat{v}_\phi}{\hat{\gamma}+1}
		\right)
	\right]
		\frac{\partial}{\partial \bar{r}}
\nonumber\\
&&
	+\left[
		\left(
			\frac{\hat{\gamma}^2\hat{v}_r\hat{v}_\theta}{\hat{\gamma}+1}
		\right)
		+\gamma_{r\phi}\sqrt{\frac{\gamma^{rr}}{\gamma_{\phi\phi}}}
		\left(
			\frac{\hat{\gamma}^2\hat{v}_\theta\hat{v}_\phi}{\hat{\gamma}+1}
		\right)
	\right]
		\frac{\partial}{\partial \bar{\theta}}
	+\left[
		\left(
			\frac{\hat{\gamma}^2\hat{v}_r\hat{v}_\phi}{\hat{\gamma}+1}
		\right)
		+\gamma_{r\phi}\sqrt{\frac{\gamma^{rr}}{\gamma_{\phi\phi}}}
		\left(
			1+
			\frac{\hat{\gamma}^2\hat{v}_\phi^2}{\hat{\gamma}+1}
		\right)
	\right]
		\frac{\partial}{\partial \bar{\phi}},\nonumber\\
\frac{1}{\sqrt{\mathstrut \gamma_{\theta\theta}}}
\frac{\partial}{\partial \theta}&=&
	-\hat{\gamma} \hat{v}_\theta \frac{\partial}{\partial \bar{t}}
	+\left(
		\frac{\hat{\gamma}^2\hat{v}_r \hat{v}_\theta}{\hat{\gamma}+1}
	\right)
		\frac{\partial}{\partial \bar{r}}
	+\left(
		1+\frac{\hat{\gamma}^2\hat{v}_\theta^2}{\hat{\gamma}+1}
	\right)
		\frac{\partial}{\partial \bar{\theta}}
	+\left(
		\frac{\hat{\gamma}^2\hat{v}_\theta\hat{v}_\phi}{\hat{\gamma}+1}
	\right)
		\frac{\partial}{\partial \bar{\phi}},\nonumber\\
\frac{1}{\sqrt{\mathstrut \gamma_{\phi\phi}}}
\frac{\partial}{\partial \phi}&=&
	-\hat{\gamma} \hat{v}_\phi \frac{\partial}{\partial \bar{t}}
	+\left(
		\frac{\hat{\gamma}^2 \hat{v}_r \hat{v}_\phi}{\hat{\gamma}+1}
	\right)
		\frac{\partial}{\partial \bar{r}}
	+\left(
		\frac{\hat{\gamma}^2 \hat{v}_\theta \hat{v}_\phi}{\hat{\gamma}+1}
	\right)
		\frac{\partial}{\partial \bar{\theta}}
	+\left(
		1+\frac{\hat{\gamma}^2\hat{v}_\phi^2}{\hat{\gamma}+1}
	\right)
		\frac{\partial}{\partial \bar{\phi}}. 
\end{eqnarray}
%%%%%
%

%%%%%%%%%%%%%%%%%%%%%%%%%%%%%%%%%%%%%%%%%%%%%%%%%%%%%%%%%%%%%%%%%%%%%%%%%%%
\subsection{Radiation moments}
The radiation energy density $E$, the radiation flux $F^i$ and 
the radiation pressure tensor $P^{ij}$ are defined as the zeroth, 
the first and the second moments of the specific intensity 
$I_\nu(x^\mu,~{\bf n})$, respectively. 
We denote the radiation moments as 
%
%%%%%
%\begin{equation}
%     E=\int\int I_\nu d\nu d\Omega,~~~
%   F^i=\int\int I_\nu n^i d\nu d\Omega,~~~
%P^{ij}=\int\int I_\nu n^i n^j d\nu d\Omega,
%\end{equation}
%%%%%
%
%when measured in the Boyer-Lindquist frame, 
%
%%%%%
\begin{equation}
     \hat{E}=\int\int \hat{I}_{\hat{\nu}} d\hat{\nu} d\hat{\Omega},~~~
 \hat{F}^{i}=\int\int \hat{I}_{\hat{\nu}} 
				\hat{n}^i d\hat{\nu} d\hat{\Omega},~~~
\hat{P}^{ij}=\int\int \hat{I}_{\hat{\nu}} 
				\hat{n}^i \hat{n}^j d\hat{\nu} d\hat{\Omega},
\end{equation}
%%%%%
when measured in the LNRF, 
and  
%
%%%%%
\begin{equation}
     \bar{E}=\int\int \bar{I}_{\bar{\nu}} d\bar{\nu} d\bar{\Omega},~~~
 \bar{F}^{i}=\int\int \bar{I}_{\bar{\nu}} 
				\bar{n}^i d\bar{\nu} d\bar{\Omega},~~~
\bar{P}^{ij}=\int\int \bar{I}_{\bar{\nu}} 
				\bar{n}^i \bar{n}^j d\bar{\nu} d\bar{\Omega}, 
\end{equation}
%%%%%
when measured in the comoving frame. 
Correspondingly, the radiation stress tensors 
for 
%the Boyer-Lindquist frame, 
the LNRF and the comoving frame are given as 
%
%%%%%
\begin{equation}
R^{\hat{\alpha}\hat{\beta}}=
	\left(
	\begin{array}{cccc}
	\hat{E} & \hat{F}^r 
		& \hat{F}^\theta & \hat{F}^\phi \\
	\hat{F}^r & \hat{P}^{rr} 
		& \hat{P}^{r\theta} & \hat{P}^{r\phi} \\
	\hat{F}^\theta & \hat{P}^{r\theta} 
		& \hat{P}^{\theta\theta} & \hat{P}^{\theta\phi} \\
	\hat{F}^\phi & \hat{P}^{r\phi} 
		& \hat{P}^{\theta\phi} & \hat{P}^{\phi\phi} 	
	\end{array}
	\right)
~~~~~{\rm and }~~~~~
R^{\bar{\alpha}\bar{\beta}}=
	\left(
	\begin{array}{cccc}
	\bar{E} & \bar{F}^r 
		& \bar{F}^\theta & \bar{F}^\phi \\
	\bar{F}^r & \bar{P}^{rr} 
		& \bar{P}^{r\theta} & \bar{P}^{r\phi} \\
	\bar{F}^\theta & \bar{P}^{r\theta} 
		& \bar{P}^{\theta\theta} & \bar{P}^{\theta\phi} \\
	\bar{F}^\phi & \bar{P}^{r\phi} 
		& \bar{P}^{\theta\phi} & \bar{P}^{\phi\phi} 	
	\end{array}
	\right),
\end{equation}
%%%%%
%
%and
%
%%%%%
%\begin{equation}
%R^{\bar{\alpha}\bar{\beta}}=
%	\left(
%	\begin{array}{cccc}
%	\bar{E} & \bar{F}^r 
%		& \bar{F}^\theta & \bar{F}^\phi \\
%	\bar{F}^r & \bar{P}^{rr} 
%		& \bar{P}^{r\theta} & \bar{P}^{r\phi} \\
%	\bar{F}^\theta & \bar{P}^{r\theta} 
%		& \bar{P}^{\theta\theta} & \bar{P}^{\theta\phi} \\
%	\bar{F}^\phi & \bar{P}^{r\phi} 
%		& \bar{P}^{\theta\phi} & \bar{P}^{\phi\phi} 	
%	\end{array}
%	\right),
%\end{equation}
%%%%%
%
respectively. 
The contravariant components of the radiation stress tensor 
$R^{\alpha\beta}$ are calculated from $R^{\hat{\alpha}\hat{\beta}}$ 
by the transformation as 
%
%%%%%
\begin{equation}
R^{\alpha\beta}=
	\frac{\partial x^\alpha}{\partial x^{\hat{\mu}}}
	\frac{\partial x^\beta}{\partial x^{\hat{\nu}}}
	R^{\hat{\mu}\hat{\nu}},  
\end{equation}
%%%%%
%
and explicitly given as 
%
%%%%%
\begin{eqnarray}
R^{\alpha\beta}&=&
\left[
	\begin{array}{cc}
	\displaystyle 
	\frac{\hat{E}}{\alpha^2} 
		& \displaystyle 
		\frac{1}{\alpha}
		\left(
			\sqrt{\mathstrut \gamma^{rr}}\hat{F}^r
			-\frac{\beta^r}{\alpha}\hat{E}
		\right)
		\vspace{.5em}
		\\
	\displaystyle 
		\frac{1}{\alpha}
		\left(
			\sqrt{\mathstrut \gamma^{rr}}\hat{F}^r
			-\frac{\beta^r}{\alpha}\hat{E}
		\right)
		& \displaystyle 
			\gamma^{rr} \hat{P}^{rr}
			-\frac{2\beta^r}{\alpha}\sqrt{\mathstrut \gamma^{rr}}\hat{F}^r
			+\left(\frac{\beta^r}{\alpha}\right)^2 \hat{E}
		\vspace{.5em}
		\\
	\displaystyle 
	\frac{\hat{F}^\theta}{\alpha\sqrt{\mathstrut \gamma_{\theta\theta}}}
		& \displaystyle 
			\frac{1}{\sqrt{\mathstrut \gamma_{\theta\theta}}}
			\left(
				\sqrt{\mathstrut \gamma^{rr}}\hat{P}^{r\theta}
				-\frac{\beta^r}{\alpha}\hat{F}^\theta
			\right)
		\vspace{.5em}
		\\
	\displaystyle 
			\frac{1}{\alpha}
			\left(
				\frac{\gamma^{r\phi}}{\sqrt{\mathstrut\gamma^{rr}}}
				\hat{F}^r
				+\frac{\hat{F}^\phi}{\sqrt{\mathstrut \gamma_{\phi\phi}}}
			\right)
		& \displaystyle 
			\sqrt{\mathstrut\gamma^{rr}}
			\left(
				\frac{\hat{P}^{r\phi}}{\sqrt{\mathstrut\gamma_{\phi\phi}}}
				+\frac{\gamma^{r\phi}}{\sqrt{\mathstrut\gamma^{rr}}}
					\hat{P}^{rr}
			\right)
			-\frac{\beta^r}{\alpha}
			\left(
				\frac{\gamma^{r\phi}}{\sqrt{\mathstrut\gamma^{rr}}}
				\hat{F}^r
				+\frac{\hat{F}^\phi}{\sqrt{\mathstrut \gamma_{\phi\phi}}}
			\right)
\end{array}
\right. \nonumber\\
&& ~~~~~ ~~~~~ ~~~~~ 
\left.
	\begin{array}{cc}
	\displaystyle 
			\frac{\hat{F}^\theta}
					{\alpha\sqrt{\mathstrut \gamma_{\theta\theta}}}
		& \displaystyle 
			\frac{1}{\alpha}
			\left(
				\frac{\gamma^{r\phi}}{\sqrt{\mathstrut\gamma^{rr}}}
				\hat{F}^r
				+\frac{\hat{F}^\phi}{\sqrt{\mathstrut \gamma_{\phi\phi}}}
			\right)
		\vspace{.5em}
		\\
	\displaystyle
			\frac{1}{\sqrt{\mathstrut \gamma_{\theta\theta}}}
			\left(
				\sqrt{\mathstrut \gamma^{rr}}\hat{P}^{r\theta}
				-\frac{\beta^r}{\alpha}\hat{F}^\theta
			\right)
		& \displaystyle 
			\sqrt{\mathstrut\gamma^{rr}}
			\left(
				\frac{\hat{P}^{r\phi}}{\sqrt{\mathstrut\gamma_{\phi\phi}}}
				+\frac{\gamma^{r\phi}}{\sqrt{\mathstrut\gamma^{rr}}}
					\hat{P}^{rr}
			\right)
			-\frac{\beta^r}{\alpha}
			\left(
				\frac{\gamma^{r\phi}}{\sqrt{\mathstrut\gamma^{rr}}}
				\hat{F}^r
				+\frac{\hat{F}^\phi}{\sqrt{\mathstrut \gamma_{\phi\phi}}}
			\right)
		\vspace{.5em}
		\\
	\displaystyle 
			\frac{\hat{P}^{\theta\theta}}{\gamma_{\theta\theta}}
		& \displaystyle 
			\frac{1}{\sqrt{\mathstrut \gamma_{\theta\theta}}}
			\left(
				\frac{\hat{P}^{\theta\phi}}
						{\sqrt{\mathstrut \gamma_{\phi\phi}}}
				+\frac{\gamma^{r\phi}}{\sqrt{\mathstrut\gamma^{rr}}}
					\hat{P}^{r\theta}
			\right)
		\vspace{.5em}
		\\
	\displaystyle 
			\frac{1}{\sqrt{\mathstrut \gamma_{\theta\theta}}}
			\left(
				\frac{\hat{P}^{\theta\phi}}
						{\sqrt{\mathstrut \gamma_{\phi\phi}}}
				+\frac{\gamma^{r\phi}}{\sqrt{\mathstrut\gamma^{rr}}}
					\hat{P}^{r\theta}
			\right)
		& \displaystyle 
			\frac{\hat{P}^{\phi\phi}}{\gamma_{\phi\phi}}
			+\frac{2\gamma^{r\phi}}
				{\sqrt{\mathstrut \gamma^{rr}\gamma_{\phi\phi}}}
				\hat{P}^{r\phi}
			+\frac{(\gamma^{r\phi})^2}{\gamma^{rr}} 
			\hat{P}^{rr}
\end{array}
\right]. 
\label{eq:RmomCFbyLNRF}
\end{eqnarray}
%%%%%
%
By using the Lorentz transformations, 
the radiation moments measured in the comoving frame 
are calculated from those measured in the LNRF as 
\begin{equation}
R^{\bar{\alpha}\bar{\beta}}=
	\Lambda^{\bar{\alpha}}_{~\hat{\mu}}(-{\rm v})
	\Lambda^{\bar{\beta}}_{~\hat{\nu}}(-{\rm v})
	R^{\hat{\mu}\hat{\nu}},  
\end{equation}
and explicitly given as \citep{MM84,P06}
%
%%%%%
\begin{eqnarray}
\bar{E} &=& \hat{\gamma}^2 
		\left(
			\hat{E}-2\hat{v}_i \hat{F}^i +\hat{v}_i \hat{v}_j \hat{P}^{ij} 
		\right),\\
\bar{F}^i &=& -\hat{\gamma}^2 \hat{v}^i \hat{E} 
		+\hat{\gamma}\left[
			\delta^i_j+\left(\hat{\gamma}+\frac{\hat{\gamma}^2}
			{\hat{\gamma}+1}\right)\hat{v}^i \hat{v}_j
		\right]\hat{F}^j
		-\hat{\gamma} \hat{v}_j \left(
			\delta^i_k+\frac{\hat{\gamma}^2}{\hat{\gamma}+1}
			\hat{v}^i \hat{v}_k
		\right) \hat{P}^{jk},\\
\bar{P}^{ij} &=& \hat{\gamma}^2 \hat{v}^i \hat{v}^j \hat{E}
		-\hat{\gamma} \left(
			\hat{v}^i \delta^j_k +\hat{v}^j \delta^i_k 
			+2\frac{\hat{\gamma}^2}{\hat{\gamma}+1} 
			\hat{v}^i \hat{v}^j \hat{v}_k
		\right)\hat{F}^k
		+\left(
			\delta^i_k+\frac{\hat{\gamma}^2}{\hat{\gamma}+1}
			\hat{v}^i \hat{v}_k
		\right)
		\left(
			\delta^j_l+\frac{\hat{\gamma}^2}{\hat{\gamma}+1}
			\hat{v}^j \hat{v}_l
		\right)
		\hat{P}^{kl}. 
\end{eqnarray}
%%%%%
%
Inversely, the radiation moments measured in the LNRF 
are calculated from those measured in the comoving frame as 
\begin{equation}
R^{\hat{\alpha}\hat{\beta}}=
	\Lambda^{\hat{\alpha}}_{~\bar{\mu}}({\rm v})
	\Lambda^{\hat{\beta}}_{~\bar{\nu}}({\rm v})
	R^{\bar{\mu}\bar{\nu}},  
\end{equation}
and explicitly given as 
%
%%%%%
\begin{eqnarray}
\hat{E} &=& \hat{\gamma}^2 
		\left(
			\bar{E}+2\hat{v}_i \bar{F}^i +\hat{v}_i \hat{v}_j \bar{P}^{ij} 
		\right),\\
\hat{F}^i &=& \hat{\gamma}^2 \hat{v}^i \bar{E} 
		+\hat{\gamma}\left[
			\delta^i_j+\left(\hat{\gamma}
			+\frac{\hat{\gamma}^2}{\hat{\gamma}+1}\right)\hat{v}^i \hat{v}_j
		\right]\bar{F}^j
		+\hat{\gamma} \hat{v}_j \left(
			\delta^i_k+\frac{\hat{\gamma}^2}{\hat{\gamma}+1}
			\hat{v}^i \hat{v}_k
		\right) \bar{P}^{jk},\\
\hat{P}^{ij} &=& \hat{\gamma}^2 \hat{v}^i \hat{v}^j \bar{E}
		+\hat{\gamma} \left(
			\hat{v}^i \delta^j_k +\hat{v}^j \delta^i_k 
			+2\frac{\hat{\gamma}^2}{\hat{\gamma}+1} 
			\hat{v}^i \hat{v}^j \hat{v}_k
		\right)\bar{F}^k
		+\left(
			\delta^i_k+\frac{\hat{\gamma}^2}{\hat{\gamma}+1}
			\hat{v}^i \hat{v}_k
		\right)
		\left(
			\delta^j_l+\frac{\hat{\gamma}^2}{\hat{\gamma}+1}
			\hat{v}^j \hat{v}_l
		\right)
		\bar{P}^{kl}. 
\end{eqnarray}
%%%%%
%

%%%%%%%%%%%%%%%%%%%%%%%%%%%%%%%%%%%%%%%%%%%%%%%%%%%%%%%%%%%%%%%%%%%%%%%%%%%%
\subsection{Radiation four-force density}
The radiation four-force density measured in the comoving frame is 
given as \citep{MM84}
%
%%%%%
\begin{equation}
G^{\bar{\alpha}}=\frac{1}{c}\int \int~
	(\bar{\chi}\bar{I}_{\bar{\nu}}-\bar{\eta})n^{\bar{\alpha}}~
	d\bar{\nu}d\bar{\Omega}, 
\end{equation}
%%%%%
%
where $\bar{\chi}$ and $\bar{\eta}$ are the mean opacity and the emissivity 
measured in the comoving frame, respectively. 
The time component $G^{\bar{t}}$
has the dimension $c^{-1}$ times the net rate of the 
radiation energy per unit volume, and 
the spatial component $G^{\bar{i}}$ has the dimension of the net rate 
of the momentum exchange between the matter and the radiation. 
The time component of the radiation four-force density 
measured in the comoving frame can be calculated as \citep{MM84,P06}
%
%%%%%
\begin{equation}
G^{\bar{t}}=\bar{\Gamma}-\bar{\Lambda}
\end{equation}
%%%%%
%
where the heating function $\bar{\Gamma}$ and the cooling function 
$\bar{\Lambda}$ are defined as 
%
%%%%%
\begin{equation}
\bar{\Gamma}\equiv\frac{1}{c}\int\int~ \bar{\chi}\bar{I}_{\bar{\nu}} 
			d\bar{\nu}d\bar{\Omega},~~~~~
\bar{\Lambda}\equiv\frac{1}{c}\int\int~ \bar{\eta} 
			d\bar{\nu}d\bar{\Omega}.
\end{equation}
%%%%%
%
The components of the radiation force $G^\alpha$ is calculated from 
those in the comoving frame $G^{\bar{\alpha}}$ by the transformation 
%
%%%%%
\begin{equation}
G^\alpha=\frac{\partial x^\alpha}{\partial x^{\bar{\beta}}}G^{\bar{\beta}}, 
\end{equation}
%%%%%
%
and explicitly given as 
%
%%%%%
\begin{eqnarray}
G^t &=& \frac{\hat{\gamma}}{\alpha}
		\left(
			G^{\bar{t}}+\hat{v}_i G^{\bar{i}}
		\right),\nonumber\\
G^r &=& \sqrt{\mathstrut\gamma^{rr}} G^{\bar{r}}
			+\hat{\gamma}
			\left(
				\sqrt{\mathstrut\gamma^{rr}}\hat{v}_r
				-\frac{\beta^r}{\alpha}
			\right)G^{\bar{t}}
			+\hat{\gamma}
			\left(\sqrt{\mathstrut\gamma^{rr}}
				\frac{\hat{\gamma}\hat{v}_r}{\hat{\gamma}+1}
				-\frac{\beta^r}{\alpha}
			\right)\hat{v}_i G^{\bar{i}}
		,\nonumber\\
G^\theta &=& \frac{1}{\sqrt{\mathstrut \gamma_{\theta\theta}}}
		\left(
			G^{\bar{\theta}}
			+\hat{\gamma}\hat{v}_\theta G^{\bar{t}}
			+\frac{\hat{\gamma}^2\hat{v}_\theta}{\hat{\gamma}+1}
				\hat{v}_i G^{\bar{i}}
		\right),\nonumber\\
G^\phi &=& 
		\frac{G^{\bar{\phi}}}{\sqrt{\mathstrut\gamma_{\phi\phi}}}
		+\frac{\gamma^{r\phi}}{\sqrt{\mathstrut\gamma^{rr}}}G^{\bar{r}}
		+\hat{\gamma}
		\left(
			\frac{\hat{v}_\phi}{\sqrt{\mathstrut \gamma_{\phi\phi}}}
			+\frac{\gamma^{r\phi}}{\sqrt{\mathstrut\gamma^{rr}}}\hat{v}_r
		\right)
		\left(
			G^{\bar{t}}
			+\frac{\hat{\gamma}}{\hat{\gamma}+1}
			\hat{v}_i G^{\bar{i}}
		\right). 
\end{eqnarray}
%%%%%
%

%%%%%%%%%%%%%%%%%%%%%%%%%%%%%%%%%%%%%%%%%%%%%%%%%%%%%%%%%%%%%%%%%%%%%%%%%%%%
\section{Radiation hydrodynamic equations}
%

%%%%%%%%%%%%%%%%%%%%%%%%%%%%%%%%%%%%%%%%%%%%%%%%%%%%%%%%%%%%%%%%%%%%%%%%%%%%
\subsection{Continuity equation}
As a continuity equation, now we consider the particle number conservation, 
$(nu^\alpha)_{;\alpha}=0$. 
This equation can be calculated as 
%
%%%%%
\begin{equation}
%\partial_t n_* +\partial_i (n_* \hat{v}^i)=0, 
%\frac{\partial}{\partial t} n_* 
%+\frac{\partial}{\partial x^i} \left(n_* v^i\right)=0, 
\frac{\partial}{\partial t}\left( \alpha \sqrt{\mathstrut \gamma} n u^t 
					\right) 
+\frac{\partial}{\partial x^i} \left( \alpha \sqrt{\mathstrut \gamma} n u^i 
					\right)=0, 
\end{equation}
%%%%%
%
where 
%$n_*\equiv n\alpha u^t\gamma^{1/2}$, $v^i=u^i/u^t$ and  
$\gamma\equiv {\rm det}\gamma_{ij}$. 
In the case of the Kerr metric written by the Boyer-Lindquist coordinate
$\gamma=\Sigma(\Sigma+2mr)\sin^2\theta$. 
%

%%%%%%%%%%%%%%%%%%%%%%%%%%%%%%%%%%%%%%%%%%%%%%%%%%%%%%%%%%%%%%%%%%%%%%%%%%%%
\subsection{Hydrodynamic equations}
The relativistic Euler equations are obtained by the projection of the 
equation of the energy momentum conservation 
$\nabla_\beta T^{\alpha\beta}=G^\alpha$ on the specific directions 
by using the projection tensor 
$P^{\alpha\beta}=g^{\alpha\beta}+u^\alpha u^\beta$ as 
$P^\alpha_\beta \nabla_\delta T^{\beta\delta}=P^\alpha_\beta G^\beta$.    
From this, we can obtain 
$\rho_0 h_{\rm g} u^\beta \nabla_\beta u^\alpha 
+(g^{\alpha\beta}+u^\alpha u^\beta)\partial_\beta P_{g}
=G^\alpha+u^\alpha u_\beta G^\beta$. 
The Euler equations in $r$, $\theta$ and $\phi$-directions are given as  
%
%%%%%
\begin{eqnarray}
 \rho_0 h_g u^t \frac{du^r}{dt} 
+\rho_0 h_g u^i \frac{du^r}{dx^i}
+\rho_0 h_g \Gamma^r_{\beta\gamma}u^\beta u^\gamma
+(g^{tr}+u^r u^t)\frac{dP_g}{dt}
+(g^{ri}+u^r u^i)\frac{dP_g}{dx^i}
&=&
 G^r 
+u^r u_\beta G^\beta
,
\label{eq:rEuler}
\\
 \rho_0 h_g u^t \frac{du^\theta}{dt} 
+\rho_0 h_g u^i \frac{du^\theta}{dx^i}
+\rho_0 h_g \Gamma^\theta_{\beta\gamma}u^\beta u^\gamma
+u^\theta u^t\frac{dP_g}{dt}
+(g^{\theta \theta}+u^r u^\theta)\frac{dP_g}{d\theta}
&=&
 G^\theta 
+u^\theta u_\beta G^\beta
,
\label{eq:thetaEuler}
\\
 \rho_0 h_g u^t \frac{du^\phi}{dt} 
+\rho_0 h_g u^i \frac{du^\phi}{dx^i}
+\rho_0 h_g \Gamma^\phi_{\beta\gamma}u^\beta u^\gamma
+(g^{t\phi}+u^r u^t)\frac{dP_g}{dt}
+(g^{\phi i}+u^r u^i)\frac{dP_g}{dx^i}
&=&
 G^\phi 
+u^\phi u_\beta G^\beta
,
\label{eq:phiEuler}
\end{eqnarray}
%%%%%
%
where
% 
%%%%%
\begin{eqnarray}
\Gamma^\alpha_{\beta\gamma}u^\beta u^\gamma
&=&\Gamma^\alpha_{tt} (u^t)^2 + \Gamma^\alpha_{rr} (u^r)^2 
+\Gamma^\alpha_{\theta\theta} (u^\theta)^2 
+\Gamma^\alpha_{\phi\phi} (u^\phi)^2 + 
\nonumber\\
&&
+2\Gamma^\alpha_{tr} u^t u^r
+2\Gamma^\alpha_{t\theta} u^t u^\theta
+2\Gamma^\alpha_{t\phi} u^t u^\phi
+2\Gamma^\alpha_{r\theta} u^r u^\theta 
+2\Gamma^\alpha_{r\phi} u^r u^\phi
+2\Gamma^\alpha_{\theta\phi} u^\theta u^\phi,       
\end{eqnarray}
%%%%%
%
and now $\alpha=r,~\theta,~\phi$. 
The Christoffel symbols are given in App. \ref{app:metric}.  
$u_\alpha G^\alpha$ is calculated 
by the heating and cooling function defined in the comoving frame 
as 
%
%%%%%
\begin{equation}
u_\alpha G^\alpha=-G^{\bar{t}}=\bar{\Lambda}-\bar{\Gamma}.  
\end{equation}
%%%%%
%
This is also calculated as 
%
%%%%%
\begin{eqnarray}
u_\alpha G^\alpha &=&  
%	\left\{
%		(-\alpha+\beta^r \beta_r)G^t 
%		+\gamma_{\phi\phi}\Omega G^\phi
%		+\gamma_{r\phi}
%			\left[
%				\Omega G^r +\beta^r (\Omega G^t +G^r +G^\phi)
%			\right]
%	\right\}
%	u^t
%	+\left[
%		\gamma_{rr}(G^r+\beta^r G^t)+\gamma_{r\phi}G^\phi
%	\right]
%	u^r
%	+\gamma_{\theta\theta}G^\theta u^\theta.
	\left[
		(-\alpha+\beta^r \beta_r)G^t 
		+\gamma_{r\phi}
				\beta^r (G^r +G^\phi)
	\right]
	u^t
	+\left[
		\gamma_{rr}(G^r+\beta^r G^t)+\gamma_{r\phi}G^\phi
	\right]
	u^r
	+\gamma_{\theta\theta}G^\theta u^\theta
	\nonumber\\
	&&
	+\left[
		\gamma_{\phi\phi} G^\phi
		+\gamma_{r\phi}(G^r +\beta^r G^t)
	\right]
	u^\phi.
\end{eqnarray}
%%%%%
%
The he local energy conservation is obtained from 
$u_\alpha \nabla_\beta T^{\alpha\beta}=u_\alpha G^\alpha$ calculated as 
%
%%%%%
\begin{eqnarray}
-n u^t \frac{\partial}{\partial t}\left(\frac{\rho_0 h_{\rm g}}{n}\right)
-n u^i \frac{\partial}{\partial x^i}\left(\frac{\rho_0 h_{\rm g}}{n}\right)
+u^t \frac{\partial P_g}{\partial t}
+u^i \frac{\partial P_g}{\partial x^i}
=u_\beta G^\beta. 
\label{eq:energyEq}
\end{eqnarray}
%%%%%
%
%The right hand side of Eq. (\ref{eq:energyEq}) is also calculated 
%by the heating and cooling function defined in the comoving frame 
%as 
%
%%%%%
%\begin{equation}
%u_\alpha G^\alpha=-G^{\bar{t}}=\bar{\Lambda}-\bar{\Gamma}. 
%\end{equation}
%%%%%
%

%%%%%%%%%%%%%%%%%%%%%%%%%%%%%%%%%%%%%%%%%%%%%%%%%%%%%%%%%%%%%%%%%%%%%%%%%%%%
\subsection{Radiation moment equations}
The radiation moment equation $\nabla_\beta R^{\alpha\beta}=-G^\alpha$ gives 
the equation for the energy density, the radiation flux and 
the radiation pressure tensor. 
The radiation energy equation is obtained from $t$-component of 
this equation $\nabla_\alpha R^{t\alpha}=-G^t$ calculated as 
%
%%%%%
\begin{eqnarray}
&&
\frac{\partial R^{tt}}{\partial t}
+\frac{1}{\alpha \sqrt{\gamma}}
	\left[
		\frac{\partial}{\partial r}
			\left(\alpha \sqrt{\mathstrut \gamma}~R^{tr}\right)
		+\frac{\partial}{\partial \theta}
			\left(\alpha \sqrt{\mathstrut \gamma}~R^{t\theta}\right)
	\right]
+\frac{\partial R^{t\phi}}{\partial \phi}
+\Gamma^t_{\beta\gamma}R^{\beta\gamma}
=-G^t. 
\label{eq:RME0}
\end{eqnarray}
%%%%%
%
The radiation momentum equation in $r$-direction 
$\nabla_\alpha R^{r\alpha}=-G^r$ is calculated as 
%
%%%%%
\begin{eqnarray}
&&
\frac{\partial R^{tr}}{\partial t}
+\frac{1}{\alpha \sqrt{\gamma}}
	\left[
		\frac{\partial}{\partial r}
			\left(\alpha \sqrt{\mathstrut \gamma}~R^{rr}\right)
		+\frac{\partial}{\partial \theta}
			\left(\alpha \sqrt{\mathstrut \gamma}~R^{r\theta}\right)
	\right]
+\frac{\partial R^{r\phi}}{\partial \phi}
+\Gamma^r_{\beta\gamma}R^{\beta\gamma}
=-G^r. 
\label{eq:RME1}
\end{eqnarray}
%%%%%
%
In the similar manner, 
the radiation momentum equation in $\theta$-direction 
$\nabla_\alpha R^{\theta\alpha}=-G^\theta$ is calculated as 
%
%%%%%
\begin{eqnarray}
&&
\frac{\partial R^{t\theta}}{\partial t}
+\frac{1}{\alpha \sqrt{\gamma}}
	\left[
		\frac{\partial}{\partial r}
			\left(\alpha \sqrt{\mathstrut \gamma}~R^{r\theta}\right)
		+\frac{\partial}{\partial \theta}
			\left(\alpha \sqrt{\mathstrut \gamma}~R^{\theta\theta}\right)
	\right]
+\frac{\partial R^{\theta\phi}}{\partial \phi}
+\Gamma^\theta_{\beta\gamma}R^{\beta\gamma}
=-G^\theta. 
\label{eq:RME2}
\end{eqnarray}
%%%%%
%
The radiation momentum equation in $\phi$-direction 
$\nabla_\alpha R^{\phi\alpha}=-G^\phi$ is calculated as 
%
%%%%%
\begin{eqnarray}
&&
\frac{\partial R^{t\phi}}{\partial t}
+\frac{1}{\alpha \sqrt{\gamma}}
	\left[
		\frac{\partial}{\partial r}
			\left(\alpha \sqrt{\mathstrut \gamma}~R^{r\phi}\right)
		+\frac{\partial}{\partial \theta}
			\left(\alpha \sqrt{\mathstrut \gamma}~R^{\theta\phi}\right)
	\right]
+\frac{\partial R^{\phi\phi}}{\partial \phi}
+\Gamma^\phi_{\beta\gamma}R^{\beta\gamma}
=-G^\phi. 
\label{eq:RME3}
\end{eqnarray}
%%%%%
%
%
Finally, 
by inserting the components of the radiation stress tensor given by 
Eq. (\ref{eq:RmomCFbyLNRF}) into Eqs. (\ref{eq:RME0}), (\ref{eq:RME1}), 
(\ref{eq:RME2}) and (\ref{eq:RME3}), we obtain the radiation moment 
equations as 
%
%%%%%
\begin{eqnarray}
&&
\frac{\partial}{\partial t} 
	\left(\frac{\hat{E}}{\alpha^2}\right) % R^{tt}
+\frac{1}{\alpha \sqrt{\gamma}}
	%\left\{
		\frac{\partial}{\partial r}
			\left[
				\sqrt{\mathstrut \gamma}~
				%R^{tr}
				%\frac{1}{\alpha}
					\left(
					\sqrt{\mathstrut \gamma^{rr}}\hat{F}^r
					-\frac{\beta^r}{\alpha}\hat{E}
					\right)
			\right]
+\frac{1}{\alpha \sqrt{\gamma}} \frac{\partial}{\partial \theta}
			\left(
				%\sqrt{\mathstrut \gamma}~%R^{t\theta}
				%\left(
					\frac{\sqrt{\mathstrut \gamma}\hat{F}^\theta}
					{\sqrt{\mathstrut \gamma_{\theta\theta}}}
				%\right)
			\right)
	%\right\}
+\frac{\partial}{\partial \phi}
	%R^{t\phi}
	\left[
			\frac{1}{\alpha}
			\left(
				\frac{\gamma^{r\phi}}{\sqrt{\mathstrut\gamma^{rr}}}
				\hat{F}^r
				+\frac{\hat{F}^\phi}{\sqrt{\mathstrut \gamma_{\phi\phi}}}
			\right)
	\right]
\nonumber\\
&&
%-\frac{1}{2}g^{tr}\partial_r g_{tt} %R^{tt}
+\Gamma^t_{tt}
	\left(\frac{\hat{E}}{\alpha^2}\right) % R^{tt}
+2\Gamma^t_{tr}
	%g^{tt}\partial_r g_{tt} 	
				%R^{tr}
				\frac{1}{\alpha}
					\left(
					\sqrt{\mathstrut \gamma^{rr}}\hat{F}^r
					-\frac{\beta^r}{\alpha}\hat{E}
					\right)
+2\Gamma^t_{t\theta}
%	 (	g^{tt}\partial_\theta g_{tt} +
%		g^{tr}\partial_\theta g_{tr}
%	)%R^{t\theta} 
				\left(
					\frac{\hat{F}^\theta}
					{\alpha\sqrt{\mathstrut \gamma_{\theta\theta}}}
				\right)
%\nonumber\\
%&&
+2\Gamma^t_{t\phi}
%	-
%	g^{tr}\partial_r g_{t\phi} %R^{t\phi}
	%R^{t\phi}
	\left[
			\frac{1}{\alpha}
			\left(
				\frac{\gamma^{r\phi}}{\sqrt{\mathstrut\gamma^{rr}}}
				\hat{F}^r
				+\frac{\hat{F}^\phi}{\sqrt{\mathstrut \gamma_{\phi\phi}}}
			\right)
	\right]
\nonumber\\
&&
+\Gamma^t_{rr} 
%	\left(	g^{tt}\partial_r g_{tr} + 
%		\frac{1}{2}g^{tr}\partial_r g_{rr}
%	\right) 
	%R^{rr}
	\left[
			\gamma^{rr} \hat{P}^{rr}
			-\frac{2\beta^r}{\alpha}\sqrt{\mathstrut \gamma^{rr}}\hat{F}^r
			+\left(\frac{\beta^r}{\alpha}\right)^2 \hat{E}	
	\right]
%\nonumber\\
%&&
+\Gamma^t_{\theta\theta}
%	-\frac{1}{2}g^{tr}\partial_r g_{\theta\theta} 
		%R^{\theta\theta}
		\left(
		\frac{\hat{P}^{\theta\theta}}{\gamma_{\theta\theta}}
		\right)
+\Gamma^t_{\phi\phi}
%	-\frac{1}{2}g^{tr}\partial_r g_{\phi\phi} 
	%R^{\phi\phi}
	\left[
			\frac{\hat{P}^{\phi\phi}}{\gamma_{\phi\phi}}
			+\frac{2\gamma^{r\phi}}
				{\sqrt{\mathstrut \gamma^{rr}\gamma_{\phi\phi}}}
				\hat{P}^{r\phi}
			+\frac{(\gamma^{r\phi})^2}{\gamma^{rr}} 
			\hat{P}^{rr}	
	\right]
\nonumber\\
&&	
+2\Gamma^t_{r\theta} 
%	(	g^{tt}\partial_\theta g_{tr} + 
%		g^{tr}\partial_\theta g_{rr}
%	) 
	%R^{r\theta} 
	\left[
			\frac{1}{\sqrt{\mathstrut \gamma_{\theta\theta}}}
			\left(
				\sqrt{\mathstrut \gamma^{rr}}\hat{P}^{r\theta}
				-\frac{\beta^r}{\alpha}\hat{F}^\theta
			\right)	
	\right]
%\nonumber\\
%&&
+2\Gamma^t_{r\phi}
%	g^{tt}\partial_r g_{t\phi} 
	%R^{r\phi}
	\left[
			\sqrt{\mathstrut\gamma^{rr}}
			\left(
				\frac{\hat{P}^{r\phi}}{\sqrt{\mathstrut\gamma_{\phi\phi}}}
				+\frac{\gamma^{r\phi}}{\sqrt{\mathstrut\gamma^{rr}}}
					\hat{P}^{rr}
			\right)
			-\frac{\beta^r}{\alpha}
			\left(
				\frac{\gamma^{r\phi}}{\sqrt{\mathstrut\gamma^{rr}}}
				\hat{F}^r
				+\frac{\hat{F}^\phi}{\sqrt{\mathstrut \gamma_{\phi\phi}}}
			\right)	
	\right]
\nonumber\\
&&
+2\Gamma^t_{\theta\phi}
%	(	g^{tt}\partial_\theta g_{t\phi} + 
%		g^{tr}\partial_\theta g_{r\phi}
%	) 
	%R^{\theta\phi}
	\left[
			\frac{1}{\sqrt{\mathstrut \gamma_{\theta\theta}}}
			\left(
				\frac{\hat{P}^{\theta\phi}}
						{\sqrt{\mathstrut \gamma_{\phi\phi}}}
				+\frac{\gamma^{r\phi}}{\sqrt{\mathstrut\gamma^{rr}}}
					\hat{P}^{r\theta}
			\right)	
	\right]
%\nonumber\\
%&&
=-G^t,
%%%%%%%%%%%%%%%%%%%%%%%%%%%%%%%%%%%%%%%%%%%%%%%%%%%%%%%%%%%%%%%%%%%%%%%%%
\\
&&
\frac{\partial}{\partial t}
	\left[		%R^{tr}
				\frac{1}{\alpha}
					\left(
					\sqrt{\mathstrut \gamma^{rr}}\hat{F}^r
					-\frac{\beta^r}{\alpha}\hat{E}
					\right)
	\right]
+\frac{1}{\alpha \sqrt{\gamma}}
	%\left[
		\frac{\partial}{\partial r}
		\left\{
			\alpha \sqrt{\mathstrut \gamma}~%R^{rr}
			%R^{rr}
			\left[
			\gamma^{rr} \hat{P}^{rr}
			-\frac{2\beta^r}{\alpha}\sqrt{\mathstrut \gamma^{rr}}\hat{F}^r
			+\left(\frac{\beta^r}{\alpha}\right)^2 \hat{E}	
			\right]
		\right\}
\nonumber\\
&&
		+\frac{1}{\alpha \sqrt{\gamma}}
			\frac{\partial}{\partial \theta}
			%\left\{
			%\alpha \sqrt{\mathstrut \gamma}~%R^{r\theta}
				%R^{r\theta} 
				\left[
					\frac{\alpha \sqrt{\mathstrut \gamma}}
						{\sqrt{\mathstrut \gamma_{\theta\theta}}}
					\left(
						\sqrt{\mathstrut \gamma^{rr}}\hat{P}^{r\theta}
						-\frac{\beta^r}{\alpha}\hat{F}^\theta
					\right)	
				\right]
			%\right\}
	%\right]
+\frac{\partial}{\partial \phi}
	%R^{r\phi}
	\left[
			\sqrt{\mathstrut\gamma^{rr}}
			\left(
				\frac{\hat{P}^{r\phi}}{\sqrt{\mathstrut\gamma_{\phi\phi}}}
				+\frac{\gamma^{r\phi}}{\sqrt{\mathstrut\gamma^{rr}}}
					\hat{P}^{rr}
			\right)
			-\frac{\beta^r}{\alpha}
			\left(
				\frac{\gamma^{r\phi}}{\sqrt{\mathstrut\gamma^{rr}}}
				\hat{F}^r
				+\frac{\hat{F}^\phi}{\sqrt{\mathstrut \gamma_{\phi\phi}}}
			\right)	
	\right]
\nonumber\\
&&
+\Gamma^r_{tt}
	%-\frac{1}{2}g^{rr}\partial_r g_{tt} %R^{tt}	
	\left(\frac{\hat{E}}{\alpha^2}\right) % R^{tt}
%\nonumber\\
%&&
+2\Gamma^r_{tr}
%	(	g^{tr}\partial_r g_{tt} + 
%		g^{r\phi}\partial_r g_{t\phi}
%	)%R^{tr} 
		\left[		%R^{tr}
				\frac{1}{\alpha}
					\left(
					\sqrt{\mathstrut \gamma^{rr}}\hat{F}^r
					-\frac{\beta^r}{\alpha}\hat{E}
					\right)
	\right]	
+2\Gamma^r_{t\theta}
%	(	g^{tr}\partial_\theta g_{tt} + 
%		g^{rr}\partial_\theta g_{tr} + 
%		g^{r\phi}\partial_\theta g_{t\phi}
%	)%R^{t\theta} 
				\left(
					\frac{\hat{F}^\theta}
					{\alpha\sqrt{\mathstrut \gamma_{\theta\theta}}}
				\right)
%\nonumber\\
%&&
+2\Gamma^r_{t\phi}
%	-
%	g^{rr} \partial_r g_{t\phi} %R^{t\phi}
	%R^{t\phi}
	\left[
			\frac{1}{\alpha}
			\left(
				\frac{\gamma^{r\phi}}{\sqrt{\mathstrut\gamma^{rr}}}
				\hat{F}^r
				+\frac{\hat{F}^\phi}{\sqrt{\mathstrut \gamma_{\phi\phi}}}
			\right)
	\right]
\nonumber\\
&&
+\Gamma^r_{rr} 
%	\left(	g^{tr}\partial_r g_{tr} + 
%		\frac{1}{2}g^{rr}\partial_r g_{rr} +
%		g^{r\phi}\partial_r g_{r\phi} 
%	\right) %R^{rr} 
	%R^{rr}
	\left[
			\gamma^{rr} \hat{P}^{rr}
			-\frac{2\beta^r}{\alpha}\sqrt{\mathstrut \gamma^{rr}}\hat{F}^r
			+\left(\frac{\beta^r}{\alpha}\right)^2 \hat{E}	
	\right]
%\nonumber\\
%&&
+\Gamma^r_{\theta\theta}
%	-
%	\frac{1}{2}g^{rr}\partial_r g_{\theta\theta} %R^{\theta\theta}
		%R^{\theta\theta}
		\left(
		\frac{\hat{P}^{\theta\theta}}{\gamma_{\theta\theta}}
		\right)
+\Gamma^r_{\phi\phi}
%	-
%	\frac{1}{2}g^{rr}\partial_r g_{\phi\phi} %R^{\phi\phi}
	%R^{\phi\phi}
	\left[
			\frac{\hat{P}^{\phi\phi}}{\gamma_{\phi\phi}}
			+\frac{2\gamma^{r\phi}}
				{\sqrt{\mathstrut \gamma^{rr}\gamma_{\phi\phi}}}
				\hat{P}^{r\phi}
			+\frac{(\gamma^{r\phi})^2}{\gamma^{rr}} 
			\hat{P}^{rr}	
	\right]
\nonumber\\
&&
+2\Gamma^r_{r\theta} 
	%(	g^{tr}\partial_\theta g_{tr} + 
	%	g^{rr}\partial_\theta g_{rr} +
	%	g^{r\phi}\partial_\theta g_{r\phi} 
	%) %R^{r\theta}
	%R^{r\theta} 
	\left[
			\frac{1}{\sqrt{\mathstrut \gamma_{\theta\theta}}}
			\left(
				\sqrt{\mathstrut \gamma^{rr}}\hat{P}^{r\theta}
				-\frac{\beta^r}{\alpha}\hat{F}^\theta
			\right)	
	\right]
%\nonumber\\
%&&
+2\Gamma^r_{r\phi}
%	(	g^{tr}\partial_r g_{t\phi} +
%		g^{r\phi}\partial_r g_{\phi\phi}
%	) %R^{r\phi}
	%R^{r\phi}
	\left[
			\sqrt{\mathstrut\gamma^{rr}}
			\left(
				\frac{\hat{P}^{r\phi}}{\sqrt{\mathstrut\gamma_{\phi\phi}}}
				+\frac{\gamma^{r\phi}}{\sqrt{\mathstrut\gamma^{rr}}}
					\hat{P}^{rr}
			\right)
			-\frac{\beta^r}{\alpha}
			\left(
				\frac{\gamma^{r\phi}}{\sqrt{\mathstrut\gamma^{rr}}}
				\hat{F}^r
				+\frac{\hat{F}^\phi}{\sqrt{\mathstrut \gamma_{\phi\phi}}}
			\right)	
	\right]
\nonumber\\
&&
+2\Gamma^r_{\theta\phi}
%	(	g^{tr}\partial_\theta g_{t\phi} +
%		g^{rr}\partial_\theta g_{r\phi} +
%		g^{r\phi}\partial_\theta g_{\phi\phi}
%	) %R^{\theta\phi}
	%R^{\theta\phi}
	\left[
			\frac{1}{\sqrt{\mathstrut \gamma_{\theta\theta}}}
			\left(
				\frac{\hat{P}^{\theta\phi}}
						{\sqrt{\mathstrut \gamma_{\phi\phi}}}
				+\frac{\gamma^{r\phi}}{\sqrt{\mathstrut\gamma^{rr}}}
					\hat{P}^{r\theta}
			\right)	
	\right]
=-G^r, 
%%%%%%%%%%%%%%%%%%%%%%%%%%%%%%%%%%%%%%%%%%%%%%%%%%%%%%%%%%%%%%%%%%%%%%%%%%%
\\
&&
\frac{\partial}{\partial t} %R^{t\theta} 
				\left(
					\frac{\hat{F}^\theta}
					{\alpha\sqrt{\mathstrut \gamma_{\theta\theta}}}
				\right)
+\frac{1}{\alpha \sqrt{\gamma}}
		\frac{\partial}{\partial r}
			%\left\{
				%\alpha \sqrt{\mathstrut \gamma}~%R^{r\theta}
				%R^{r\theta} 
				\left[
					\frac{\alpha \sqrt{\mathstrut \gamma}}
						{\sqrt{\mathstrut \gamma_{\theta\theta}}}
					\left(
						\sqrt{\mathstrut \gamma^{rr}}\hat{P}^{r\theta}
						-\frac{\beta^r}{\alpha}\hat{F}^\theta
					\right)	
				\right]
			%\right\}
+\frac{1}{\alpha \sqrt{\gamma}}\frac{\partial}{\partial \theta}
			%\left[
			%\alpha \sqrt{\mathstrut \gamma}~%R^{\theta\theta}
				%R^{\theta\theta}
				\left(
				\frac{\alpha \sqrt{\mathstrut \gamma}
						\hat{P}^{\theta\theta}}{\gamma_{\theta\theta}}
				\right)
			%\right]
%\nonumber\\
%&&
+\frac{\partial}{\partial \phi}
	%R^{\theta\phi}
	\left[
			\frac{1}{\sqrt{\mathstrut \gamma_{\theta\theta}}}
			\left(
				\frac{\hat{P}^{\theta\phi}}
						{\sqrt{\mathstrut \gamma_{\phi\phi}}}
				+\frac{\gamma^{r\phi}}{\sqrt{\mathstrut\gamma^{rr}}}
					\hat{P}^{r\theta}
			\right)	
	\right]
\nonumber\\
&&
+\Gamma^\theta_{tt}
%	-\frac{1}{2}g^{\theta \theta}\partial_\theta g_{tt} %R^{tt}
		\left(\frac{\hat{E}}{\alpha^2}\right) % R^{tt}
%\nonumber\\
%&&
+\Gamma^\theta_{rr}
%	-\frac{1}{2}g^{\theta \theta}\partial_\theta g_{rr} %R^{rr}
	%R^{rr}
	\left[
			\gamma^{rr} \hat{P}^{rr}
			-\frac{2\beta^r}{\alpha}\sqrt{\mathstrut \gamma^{rr}}\hat{F}^r
			+\left(\frac{\beta^r}{\alpha}\right)^2 \hat{E}	
	\right]
%\nonumber\\
%&&
+\Gamma^\theta_{\theta\theta} 
%	\frac{1}{2} g^{\theta\theta} 
%		\partial_\theta g_{\theta\theta}%R^{\theta\theta}
		%R^{\theta\theta}
		\left(
		\frac{\hat{P}^{\theta\theta}}{\gamma_{\theta\theta}}
		\right)
%\nonumber\\
%&&
+\Gamma^\theta_{\phi\phi}
	%-\frac{1}{2}g^{\theta \theta}\partial_\theta g_{\phi\phi} %R^{\phi\phi}
	%R^{\phi\phi}
	\left[
			\frac{\hat{P}^{\phi\phi}}{\gamma_{\phi\phi}}
			+\frac{2\gamma^{r\phi}}
				{\sqrt{\mathstrut \gamma^{rr}\gamma_{\phi\phi}}}
				\hat{P}^{r\phi}
			+\frac{(\gamma^{r\phi})^2}{\gamma^{rr}} 
			\hat{P}^{rr}	
	\right]
\nonumber\\
&&
+2\Gamma^\theta_{tr}
%-g^{\theta \theta} \partial_\theta g_{tr}    %R^{tr}
	\left[		%R^{tr}
				\frac{1}{\alpha}
					\left(
					\sqrt{\mathstrut \gamma^{rr}}\hat{F}^r
					-\frac{\beta^r}{\alpha}\hat{E}
					\right)
	\right]
%\nonumber\\
%&&
+2\Gamma^\theta_{t\phi}
%-g^{\theta \theta} \partial_\theta g_{t\phi} %R^{t\phi}
	%R^{t\phi}
	\left[
			\frac{1}{\alpha}
			\left(
				\frac{\gamma^{r\phi}}{\sqrt{\mathstrut\gamma^{rr}}}
				\hat{F}^r
				+\frac{\hat{F}^\phi}{\sqrt{\mathstrut \gamma_{\phi\phi}}}
			\right)
	\right]
%\nonumber\\
%&&
+2\Gamma^\theta_{r\theta}
%g^{\theta\theta} \partial_r g_{\theta\theta} %R^{r\theta} 
	%R^{r\theta} 
	\left[
			\frac{1}{\sqrt{\mathstrut \gamma_{\theta\theta}}}
			\left(
				\sqrt{\mathstrut \gamma^{rr}}\hat{P}^{r\theta}
				-\frac{\beta^r}{\alpha}\hat{F}^\theta
			\right)	
	\right]
\nonumber\\
&&
+2\Gamma^\theta_{r\phi}
%-g^{\theta \theta} \partial_\theta g_{r\phi} %R^{r\phi}
	%R^{r\phi}
	\left[
			\sqrt{\mathstrut\gamma^{rr}}
			\left(
				\frac{\hat{P}^{r\phi}}{\sqrt{\mathstrut\gamma_{\phi\phi}}}
				+\frac{\gamma^{r\phi}}{\sqrt{\mathstrut\gamma^{rr}}}
					\hat{P}^{rr}
			\right)
			-\frac{\beta^r}{\alpha}
			\left(
				\frac{\gamma^{r\phi}}{\sqrt{\mathstrut\gamma^{rr}}}
				\hat{F}^r
				+\frac{\hat{F}^\phi}{\sqrt{\mathstrut \gamma_{\phi\phi}}}
			\right)	
	\right]
%\nonumber\\
%&&
=-G^\theta, 
%%%%%%%%%%%%%%%%%%%%%%%%%%%%%%%%%%%%%%%%%%%%%%%%%%%%%%%%%%%%%%%%%%%%%%%%%%
\\
&&
\frac{\partial}{\partial t} %R^{t\phi}
	%R^{t\phi}
	\left[
			\frac{1}{\alpha}
			\left(
				\frac{\gamma^{r\phi}}{\sqrt{\mathstrut\gamma^{rr}}}
				\hat{F}^r
				+\frac{\hat{F}^\phi}{\sqrt{\mathstrut \gamma_{\phi\phi}}}
			\right)
	\right]
+\frac{1}{\alpha \sqrt{\gamma}}
	%\left[
		\frac{\partial}{\partial r}
			\left\{
				\alpha \sqrt{\mathstrut \gamma}~%R^{r\phi}
				%R^{r\phi}
				\left[
					\sqrt{\mathstrut\gamma^{rr}}
					\left(
				\frac{\hat{P}^{r\phi}}{\sqrt{\mathstrut\gamma_{\phi\phi}}}
				+\frac{\gamma^{r\phi}}{\sqrt{\mathstrut\gamma^{rr}}}
					\hat{P}^{rr}
					\right)
					-\frac{\beta^r}{\alpha}
					\left(
						\frac{\gamma^{r\phi}}{\sqrt{\mathstrut\gamma^{rr}}}
						\hat{F}^r
					+\frac{\hat{F}^\phi}{\sqrt{\mathstrut \gamma_{\phi\phi}}}
					\right)	
				\right]
			\right\}
\nonumber\\
&&
+\frac{1}{\alpha \sqrt{\gamma}}\frac{\partial}{\partial \theta}
			%\left\{
				%\alpha \sqrt{\mathstrut \gamma}~%R^{\theta\phi}
				%R^{\theta\phi}
				\left[
					\frac{\alpha \sqrt{\mathstrut \gamma}}
							{\sqrt{\mathstrut \gamma_{\theta\theta}}}
					\left(
						\frac{\hat{P}^{\theta\phi}}
							{\sqrt{\mathstrut \gamma_{\phi\phi}}}
						+\frac{\gamma^{r\phi}}{\sqrt{\mathstrut\gamma^{rr}}}
						\hat{P}^{r\theta}
					\right)	
				\right]
			%\right\}
	%\right]
+\frac{\partial}{\partial \phi} 
	%R^{\phi\phi}
	\left[
			\frac{\hat{P}^{\phi\phi}}{\gamma_{\phi\phi}}
			+\frac{2\gamma^{r\phi}}
				{\sqrt{\mathstrut \gamma^{rr}\gamma_{\phi\phi}}}
				\hat{P}^{r\phi}
			+\frac{(\gamma^{r\phi})^2}{\gamma^{rr}} 
			\hat{P}^{rr}	
	\right]
\nonumber\\
&&
+\Gamma^\phi_{tt}
%	-\frac{1}{2}g^{r\phi}\partial_r g_{tt} %R^{tt}
		\left(\frac{\hat{E}}{\alpha^2}\right) % R^{tt}
+2\Gamma^\phi_{tr}
%	g^{\phi\phi}\partial_r g_{t\phi}
	%R^{tr}
	\left[		%R^{tr}
				\frac{1}{\alpha}
					\left(
					\sqrt{\mathstrut \gamma^{rr}}\hat{F}^r
					-\frac{\beta^r}{\alpha}\hat{E}
					\right)
	\right] 
+2\Gamma^\phi_{t\theta}
%	(	g^{r\phi}\partial_\theta g_{tr} + 
%		g^{\phi\phi}\partial_r g_{t\phi}
%	)%R^{t\theta} 
				\left(
					\frac{\hat{F}^\theta}
					{\alpha\sqrt{\mathstrut \gamma_{\theta\theta}}}
				\right)
%\nonumber\\
%&&
+2\Gamma^\phi_{t\phi}
%	-
%	g^{r\phi}\partial_r g_{t\phi} %R^{t\phi}
	%R^{t\phi}
	\left[
			\frac{1}{\alpha}
			\left(
				\frac{\gamma^{r\phi}}{\sqrt{\mathstrut\gamma^{rr}}}
				\hat{F}^r
				+\frac{\hat{F}^\phi}{\sqrt{\mathstrut \gamma_{\phi\phi}}}
			\right)
	\right]
\nonumber\\
&&
+\Gamma^\phi_{rr}
%	\left(	\frac{1}{2} g^{r\phi}\partial_r g_{rr} + 
%		g^{\phi\phi}\partial_r g_{r\phi}
%	\right) %R^{rr} 
	%R^{rr}
	\left[
			\gamma^{rr} \hat{P}^{rr}
			-\frac{2\beta^r}{\alpha}\sqrt{\mathstrut \gamma^{rr}}\hat{F}^r
			+\left(\frac{\beta^r}{\alpha}\right)^2 \hat{E}	
	\right]
%\nonumber\\
%&&
+\Gamma^\phi_{\theta\theta}
%	-\frac{1}{2}g^{r\phi}\partial_r g_{\theta\theta} %R^{\theta\theta}
		%R^{\theta\theta}
		\left(
		\frac{\hat{P}^{\theta\theta}}{\gamma_{\theta\theta}}
		\right)
+\Gamma^\phi_{\phi\phi}
%	-\frac{1}{2}g^{r\phi}\partial_r g_{\phi\phi} %R^{\phi\phi}
	%R^{\phi\phi}
	\left[
			\frac{\hat{P}^{\phi\phi}}{\gamma_{\phi\phi}}
			+\frac{2\gamma^{r\phi}}
				{\sqrt{\mathstrut \gamma^{rr}\gamma_{\phi\phi}}}
				\hat{P}^{r\phi}
			+\frac{(\gamma^{r\phi})^2}{\gamma^{rr}} 
			\hat{P}^{rr}	
	\right]
\nonumber\\
&&
+2\Gamma^\phi_{r\theta}
%	(	g^{r\phi}\partial_\theta g_{rr} + 
%		g^{\phi\phi}\partial_r g_{r\phi}
%	) %R^{r\theta} 
	%R^{r\theta} 
	\left[
			\frac{1}{\sqrt{\mathstrut \gamma_{\theta\theta}}}
			\left(
				\sqrt{\mathstrut \gamma^{rr}}\hat{P}^{r\theta}
				-\frac{\beta^r}{\alpha}\hat{F}^\theta
			\right)	
	\right]	
%\nonumber\\
%&&
+2\Gamma^\phi_{r\phi}
%	(	g^{r\phi}\partial_r g_{r\phi} +
%		g^{\phi\phi}\partial_r g_{\phi\phi}
%	) %R^{r\phi}
	%R^{r\phi}
	\left[
			\sqrt{\mathstrut\gamma^{rr}}
			\left(
				\frac{\hat{P}^{r\phi}}{\sqrt{\mathstrut\gamma_{\phi\phi}}}
				+\frac{\gamma^{r\phi}}{\sqrt{\mathstrut\gamma^{rr}}}
					\hat{P}^{rr}
			\right)
			-\frac{\beta^r}{\alpha}
			\left(
				\frac{\gamma^{r\phi}}{\sqrt{\mathstrut\gamma^{rr}}}
				\hat{F}^r
				+\frac{\hat{F}^\phi}{\sqrt{\mathstrut \gamma_{\phi\phi}}}
			\right)	
	\right] 
\nonumber\\
&&
+2\Gamma^\phi_{\theta\phi}
%	g^{\phi\phi}\partial_r g_{\phi\phi}
	%R^{\theta\phi} 
	%R^{\theta\phi}
	\left[
			\frac{1}{\sqrt{\mathstrut \gamma_{\theta\theta}}}
			\left(
				\frac{\hat{P}^{\theta\phi}}
						{\sqrt{\mathstrut \gamma_{\phi\phi}}}
				+\frac{\gamma^{r\phi}}{\sqrt{\mathstrut\gamma^{rr}}}
					\hat{P}^{r\theta}
			\right)	
	\right]
=-G^\phi,  
\end{eqnarray}
%%%%%
%
where the Christoffel symbols are given in App. \ref{app:metric}.

%%%%%%%%%%%%%%%%%%%%%%%%%%%%%%%%%%%%%%%%%%%%%%%%%%%%%%%%%%%%%%%%%%%%%%%
\section{Concluding remarks}

In this study, we have derived 
equations of fully general relativistic radiation hydrodynamics 
around a rotating black hole 
by using the Kerr-Schild coordinate 
where there is no coordinate singularity at the event horizon.  
Both the matter and the radiation are affected by the frame-dragging 
effects due to the black hole's rotation. 
Since 
the radiation usually interact with matter moving at 
relativistic velocities near the event horizon,  
the interaction between the radiation and the matter should be 
treated fully relativistically. 
This can be done if the interplay between the radiation and the 
matter is evaluated in the comoving frame where the matter is at rest. 
In this approach 
while the interactions between matter and radiation are introduced 
in the comoving frame, the equations and the derivatives 
for the description of the global evolution of both matter and the 
radiation are given in the Kerr-Schild frame (KSF) which is a frame 
fixed to the coordinate describing the central black hole. 
As a frame fixed to the coordinate, 
we use the locally non-rotating reference frame (LNRF) 
which is a radially falling frame when the Kerr-Schild coordinate 
is used. 
We can see in the derived equations that 
both the matter and the radiation are affected by the frame-dragging 
effects due to the black hole's rotation. 
It is widely known that the moment equations truncated at the finite 
order do not produce the complete system of equations, i.e., 
the number of variables are larger than the number of equations. 
So, the additional equations are required to close the system of equations
\citep{C60,M70,P73,RL79,MM84,S91}. 
These additional equations should be also described fully relativistically 
when the radiation interact with the matter moving at relativistic 
velocities. 
As one possible way with respect to such additional equations, 
the flux-limited diffusion approximation proposed by \cite{LP81} 
is sometimes used. 
The covariant theories for the flux-limited diffusion approximation 
is presented for radiation propagating through inhomogeneous and 
nonstationary media \citep{AR92}. 
It is noted that constructing the closure relations which are 
essentially required for the radiation hydrodynamical calculations 
is non-trivial when the photons and the matters are relatively 
relativistically moving. 
This is mainly because the truncation of the moment equations at the 
finite order can not be performed frame-independently. 
So, in such cases, the equations given in this articles are of no use 
if the sufficient closure relations can not be constructed. 
In the case of the calculations based on the 
Kerr-Schild coordinate, the fluid velocity is usually smaller than 
the speed of light even at the event horizon \citep{T07a}, for example, 
the gamma factor of the fluid at the horizon $\gamma < 1$-100. 
So, if the photon motion have the speed similar to the fluid's speed and 
the closure relations which is correct for such motions can be constructed, 
the equations given in this study can be used for the description of such 
radiation hydrodynamic flows. 
It is also note that when effects of the radiation stress can not be 
neglected, the dissipation effects can not be correctly treated so far. 
This is mainly because we do not know the covariant theory describing 
the local dissipation effects such as viscosity. 
One of the promising approach is given by the method based on the 
extended causal thermodynamics such as the Israel-Stewart theory 
\citep{is79} where the causality violating infinite signal speeds 
are eliminated (see also, Anile, Pav\'{o}n \& Romano 1998). 
However, while the hydrodynamical equations based on such theory are 
formulated by \cite{pa98}, there is no formulations based on such 
theory for the radiation hydrodynamic or magnetohydrodynamic equations.  
This is one of the important studies in terms of the relativistic 
radiation hydrodynamics in future. 
%

%
%\clearpage
\section*{Acknowledgments}
The author is grateful to 
Professors Y. Eriguchi and S. Mineshige for their continuous encouragements,
J. Fukue, Y. Sekiguchi, M. Shibata for useful discussion, 
and 
A. Liebmann for proofreading.
The author also thanks to Department of Physics at Montana State University 
for its hospitality and Professor S. Tsuruta for her hospitality. 
This research was partially supported by the Ministry of Education,
Culture, Sports, Science and Technology, Grant-in-Aid for 
Japan Society for the Promotion of Science (JSPS) Fellows (17010519).

%[Ref: Old Relativists' references, Euler ... ]

\appendix

%%%%%%%%%%%%%%%%%%%%%%%%%%%%%%%%%%%%%%%%%%%%%%%%%%%%%%%%%%%%%%%%%%%%%%%%%%%%
\section{Differential values of metric components 
and Christoffel symbols}
\label{app:metric}
Here, we present the explicit expressions for the metric components and 
the their differential values  
with respect to $r$ and $\theta$ used in this paper. 
Non-zero components of the metric are given as  
%
%%%%%
\begin{eqnarray}
&&
g_{tt}=-\left(1-\frac{2mr}{\Sigma}\right),~~~
g_{tr}=g_{rt}=\frac{2mr}{\Sigma},~~~
g_{t\phi}=g_{\phi t}=-\frac{2mar\sin^2\theta}{\Sigma},~~~
g_{rr}=1+\frac{2mr}{\Sigma},~~~
\nonumber\\
&&
g_{r\phi}=g_{\phi r}=-a\sin^2\theta\left(1+\frac{2mr}{\Sigma}\right),~~~
g_{\theta\theta}=\Sigma,~~~
g_{\phi\phi}=\frac{A\sin^2\theta}{\Sigma},
\end{eqnarray}
%%%%%
%
and
%
%%%%%
\begin{eqnarray}
&&
g^{tt}= -\left(1+\frac{2mr}{\Sigma}\right),~~~
g^{tr}=g^{rt}= \frac{2mr}{\Sigma},~~~
g^{rr}= \frac{\Delta}{\Sigma},~~~
g^{r\phi}=g^{\phi r}= \frac{a}{\Sigma},~~~
g^{\theta\theta}= \frac{1}{\Sigma},~~~
g^{\phi\phi} = \frac{1}{\Sigma\sin^2\theta}.
\end{eqnarray}
%%%%%
%
The differential values of the metric components 
are given as
%
%%%%%
\begin{eqnarray}
&&
\partial_r g_{tt}=\partial_r g_{tr}=\partial_r g_{rt}=\partial_r g_{rr}
=\frac{2m}{\Sigma}\left(1-\frac{2r^2}{\Sigma}\right),~~~
\partial_r g_{t\phi}=\partial_r g_{\phi t}=
\partial_r g_{r\phi}=\partial_r g_{\phi r}=
	-\frac{2ma\sin^2\theta}{\Sigma}\left(1-\frac{2r^2}{\Sigma}\right),
\nonumber\\
&&
\partial_r g_{\theta\theta}=2r,~~~
\partial_r g_{\phi\phi}
	=2\sin^2\theta
	\left[
		r + \frac{ma^2\sin^2\theta}{\Sigma}
				\left(1-\frac{2r^2}{\Sigma}\right)
	\right], 
\end{eqnarray}
%%%%%
%
and 
%
%%%%%
\begin{eqnarray}
&&
\partial_\theta g_{tt} = \partial_\theta g_{tr} = \partial_\theta g_{rt}
= \partial_\theta g_{rr}
=\frac{2ma^2r}{\Sigma^2}\sin 2\theta,~~~
\partial_\theta g_{t\phi}=\partial_\theta g_{\phi t}
=-\frac{2mar(r^2+a^2)}{\Sigma^2}\sin 2\theta,~~~
\partial_\theta g_{\theta\theta}=-a^2\sin 2\theta,
\nonumber\\
&&
\partial_\theta g_{r\phi} = \partial_\theta g_{\phi r}
=-a\sin 2\theta \left[
	1+\frac{2mr(r^2+a^2)}{\Sigma^2}
\right],~~~
\partial_\theta g_{\phi\phi}
=
\left[
\Delta + 2mr\left(\frac{r^2+a^2}{\Sigma}\right)^2
\right]\sin 2\theta
, 
\end{eqnarray}
%%%%%
%
where we have used 
%
%%%%%
\begin{eqnarray}
&&
\partial_r \Sigma = 2r,~~~
\partial_r A = 2\left[2r\Sigma+(r+m)a^2\sin^2\theta \right],~~~
\partial_\theta \Sigma = -a^2\sin 2\theta,~~~
\partial_\theta A = -a^2\Delta \sin 2\theta. 
\end{eqnarray}
%%%%%
%
The Christoffel symbols $\Gamma^\alpha_{\beta\gamma}$ 
($\alpha=t,~r,~\theta,~\phi$) are given as 
%
%%%%%
\begin{eqnarray}
&&
\Gamma^t_{tt}=-\frac{1}{2}g^{tr}\partial_r g_{tt}
=-%\frac{1}{2}
	\frac{2m^2r}{\Sigma^2} %g^{tr}
	%\frac{m}{\Sigma}
	\left(1-\frac{2r^2}{\Sigma}\right) %\partial_r g_{tt}
,~~~
%\nonumber\\
%&&
\Gamma^t_{rr}=	g^{tt}\partial_r g_{tr} + 
				\frac{1}{2}g^{tr}\partial_r g_{rr}
=	-\frac{2m}{\Sigma}\left(1+\frac{mr}{\Sigma}\right) %g^{tt}
	\left(1-\frac{2r^2}{\Sigma}\right) %\partial_r g_{tr} 
	%+ 
	%\frac{1}{2}
	%\frac{mr}{\Sigma} %g^{tr}
	%\frac{2m}{\Sigma}\left(1-\frac{2r^2}{\Sigma}\right) %\partial_r g_{rr}
,~~~
\nonumber\\
&&
\Gamma^t_{\theta\theta}=-\frac{1}{2}g^{tr}\partial_r g_{\theta\theta}
=-%\frac{1}{2}
	\frac{2mr^2}{\Sigma} %g^{tr}
	%2r%\partial_r g_{\theta\theta}
,~~~
\Gamma^t_{\phi\phi}=-\frac{1}{2}g^{tr}\partial_r g_{\phi\phi}
=-%\frac{1}{2}
	\frac{2mr\sin^2\theta}{\Sigma} %g^{tr}
	%\sin^2\theta
	\left[
		r + \frac{ma^2\sin^2\theta}{\Sigma}
				\left(1-\frac{2r^2}{\Sigma}\right)
	\right] %\partial_r g_{\phi\phi}
,~~~
\nonumber\\
&&
\Gamma^t_{tr}=
\Gamma^t_{rt}=\frac{1}{2} g^{tt}\partial_r g_{tt}
=-%\frac{1}{2} 
	\frac{m}{\Sigma}
	\left(1+\frac{2mr}{\Sigma}\right) %g^{tt}
	%\frac{2m}{\Sigma}
	\left(1-\frac{2r^2}{\Sigma}\right) %\partial_r g_{tt}
,~~~
%\nonumber\\
%&&
\Gamma^t_{t\theta}=
\Gamma^t_{\theta t}= \frac{1}{2}(g^{tt}\partial_\theta g_{tt} +
		g^{tr}\partial_\theta g_{tr})
= -%\frac{1}{2}
	%\left(1+\frac{2mr}{\Sigma}\right) %g^{tt}
	\frac{ma^2r}{\Sigma^2}\sin 2\theta %\partial_\theta g_{tt} 
	%+
	%\frac{1}{2}
	%\frac{mr}{\Sigma} %g^{tr}
	%\frac{2ma^2r}{\Sigma^2}\sin 2\theta %\partial_\theta g_{tr}
	,~~~
\nonumber\\
&&
\Gamma^t_{t\phi}=
\Gamma^t_{\phi t}=-\frac{1}{2} 
	g^{tr}\partial_r g_{t\phi}
=%\frac{1}{2} 
	%\frac{mr}{\Sigma} %g^{tr}
	\frac{2m^2ar\sin^2\theta}{\Sigma^2}\left(1-\frac{2r^2}{\Sigma}\right) 	
		%\partial_r g_{t\phi}
,~~~
%\nonumber\\
%&&
\Gamma^t_{r\theta}=
\Gamma^t_{\theta r}=\frac{1}{2} (g^{tt}\partial_\theta g_{tr} + 
		g^{tr}\partial_\theta g_{rr})
=	-%\frac{1}{2}
	%\left(1+\frac{2mr}{\Sigma}\right) %g^{tt}
	\frac{ma^2r}{\Sigma^2}\sin 2\theta %\partial_\theta g_{tr} 
	%+ 
	%\frac{1}{2}
	%\frac{mr}{\Sigma} %g^{tr}
	%\partial_\theta g_{rr}
	,~~~
\nonumber\\
&&
\Gamma^t_{r\phi}=
\Gamma^t_{\phi r}=\frac{1}{2} g^{tt}\partial_r g_{t\phi}
=%\frac{1}{2} 
	\frac{ma\sin^2\theta}{\Sigma}
	\left(1+\frac{2mr}{\Sigma}\right) %g^{tt}
	%\frac{2ma\sin^2\theta}{\Sigma}
	\left(1-\frac{2r^2}{\Sigma}\right)
	%\partial_r g_{t\phi}
,~~~
%\nonumber\\
%&&
\Gamma^t_{\theta\phi}=
\Gamma^t_{\phi\theta}=\frac{1}{2} (	g^{tt}\partial_\theta g_{t\phi} + 
		g^{tr}\partial_\theta g_{r\phi}
	)
=%\frac{1}{2} 
	%\left(1+\frac{2mr}{\Sigma}\right) %g^{tt}
	%\frac{mar(r^2+a^2)}{\Sigma^2}\sin 2\theta %\partial_\theta g_{t\phi} 
	%- 
	%\frac{1}{2} 
	%\frac{mar}{\Sigma} %g^{tr}
	%\left(
	\frac{ma^3r}{\Sigma^2}
	%\right)
	\sin^2\theta \sin 2\theta 
	%\left[
	%	1+\frac{2mr(r^2+a^2)}{\Sigma^2}
	%\right]
	%\partial_\theta g_{r\phi}
	,~~~
\nonumber\\
&&
\Gamma^r_{tt}=-\frac{1}{2}g^{rr}\partial_r g_{tt}
=-%\frac{1}{2}
	\frac{m\Delta}{\Sigma^2} %g^{rr}
	%\frac{m}{\Sigma}
	\left(1-\frac{2r^2}{\Sigma}\right) %\partial_r g_{tt}
,~~~
%\nonumber\\
%&&
\Gamma^r_{rr}=	g^{tr}\partial_r g_{tr} + 
		\frac{1}{2}g^{rr}\partial_r g_{rr} +
		g^{r\phi}\partial_r g_{r\phi} 
= \frac{m}{\Sigma}
	\left(
		1-\frac{2r^2}{\Sigma}
	\right)
	\left(
		2-\frac{\Delta}{\Sigma}
	\right)
%=	\frac{m(r^2+2mr+a^2\cos 2\theta)}{\Sigma} %g^{tr}
	%\frac{2m}{\Sigma}
%	\left(1-\frac{2r^2}{\Sigma}\right) %\partial_r g_{tr} 
	%+ 
	%\frac{1}{2}
	%\frac{\Delta}{\Sigma} %g^{rr}
	%\frac{m}{\Sigma}\left(1-\frac{2r^2}{\Sigma}\right) %\partial_r g_{rr} 
	%-
	%\frac{a}{\Sigma} %g^{r\phi}
	%\frac{2ma^2\sin^2\theta}{\Sigma^2}\left(1-\frac{2r^2}{\Sigma}\right)
	%\partial_r g_{r\phi} 
,~~~
\nonumber\\
&&
\Gamma^r_{\theta\theta}=
	-\frac{1}{2}g^{rr}\partial_r g_{\theta\theta}
=
	-%\frac{1}{2}
		\frac{r\Delta}{\Sigma} %g^{rr}
		%\partial_r g_{\theta\theta}
,~~~
\Gamma^r_{\phi\phi}=
	-\frac{1}{2}g^{rr}\partial_r g_{\phi\phi}
=
	-%\frac{1}{2}
	\frac{\Delta}{\Sigma} %g^{rr}
	\sin^2\theta
	\left[
		r + \frac{ma^2\sin^2\theta}{\Sigma}
				\left(1-\frac{2r^2}{\Sigma}\right)
	\right]
	%\partial_r g_{\phi\phi}
,~~~
\nonumber\\
&&
\Gamma^r_{tr}=
\Gamma^r_{rt}=\frac{1}{2} (	g^{tr}\partial_r g_{tt} + 
		g^{r\phi}\partial_r g_{t\phi}
	)
= \frac{m}{\Sigma}
	\left(
		1-\frac{2r^2}{\Sigma}
	\right)
	\left(
		1-\frac{\Delta}{\Sigma}
	\right)
%=%\frac{1}{2} 
	%\frac{2mr}{\Sigma} %g^{tr}
%	\frac{m(2mr-a^2\sin^2\theta)}{\Sigma^2}
%	\left(1-\frac{2r^2}{\Sigma}\right) %\partial_r g_{tt} 
	%- 
	%\frac{1}{2} 
	%\frac{a}{\Sigma} %g^{r\phi}
	%\frac{ma\sin^2\theta}{\Sigma}\left(1-\frac{2r^2}{\Sigma}\right)
	%\partial_r g_{t\phi}
,~~~
\nonumber\\
&&
\Gamma^r_{t\theta}=
\Gamma^r_{\theta t}=\frac{1}{2} (	g^{tr}\partial_\theta g_{tt} + 
		g^{rr}\partial_\theta g_{tr} + 
		g^{r\phi}\partial_\theta g_{t\phi}
	)
=0%\frac{1}{2} 
	%\frac{mr}{\Sigma} %g^{tr}
	%\frac{2m^2a^2r^2}{\Sigma^3}\sin 2\theta %\partial_\theta g_{tt} 
	%+ 
	%\frac{1}{2} 
	%\frac{\Delta}{\Sigma} %g^{rr}
	%\frac{ma^2r\Delta}{\Sigma^3}\sin 2\theta %\partial_\theta g_{tr} 
	%- 
	%\frac{1}{2} 
	%\frac{a}{\Sigma} %g^{r\phi}
	%\frac{ma^2r(r^2+a^3)}{\Sigma^2}\sin 2\theta %\partial_\theta g_{t\phi}
,~~~
%\nonumber\\
%&&
\Gamma^r_{t\phi}=
\Gamma^r_{\phi t}=-\frac{1}{2} 
	g^{rr} \partial_r g_{t\phi}
=%\frac{1}{2} 
	%\frac{\Delta}{\Sigma} %g^{rr} 
	\frac{ma\Delta\sin^2\theta}{\Sigma^2}\left(1-\frac{2r^2}{\Sigma}\right)
	%\partial_r g_{t\phi},~~~
\nonumber\\
&&
\Gamma^r_{r\theta}=
\Gamma^r_{\theta r}=\frac{1}{2} (	g^{tr}\partial_\theta g_{tr} + 
		g^{rr}\partial_\theta g_{rr} +
		g^{r\phi}\partial_\theta g_{r\phi} 
	)
=%\frac{1}{2} 
	%\frac{mr}{\Sigma} %g^{tr}
	-\frac{a^2}{2\Sigma}\sin 2\theta %\partial_\theta g_{tr} 
	%+ 
	%\frac{1}{2} 
	%\frac{\Delta}{\Sigma} %g^{rr}
	%\frac{ma^2r\Delta}{\Sigma^3}\sin 2\theta %\partial_\theta g_{rr} 
	%-
	%\frac{1}{2} 
	%\frac{a^2}{\Sigma} %g^{r\phi}
	%\sin 2\theta 
	%\left[
	%	\frac{1}{2}+\frac{mr(r^2+a^2)}{\Sigma^2}
	%\right]
	%\partial_\theta g_{r\phi} 
,~~~
\nonumber\\
&&
\Gamma^r_{r\phi}=
\Gamma^r_{\phi r}=\frac{1}{2} (	g^{tr}\partial_r g_{t\phi} +
		g^{r\phi}\partial_r g_{\phi\phi}
	)
= \frac{a}{\Sigma}\sin^2\theta
		\left[
			r - m \left(1-\frac{2r^2}{\Sigma}\right)
				\left(1-\frac{\Delta}{\Sigma}\right)
		\right]
%=%-%\frac{1}{2} 
	%\frac{mr}{\Sigma} %g^{tr}
	%\frac{2ma\sin^2\theta}{\Sigma}\left(1-\frac{2r^2}{\Sigma}\right)
	%\partial_r g_{t\phi} 
	%+
	%\frac{1}{2} 
%	\frac{a}{\Sigma} %g^{r\phi}
%	\sin^2\theta
%	\left[
%		r - \frac{m(2mr-a^2\sin^2\theta)}{\Sigma}
%				\left(1-\frac{2r^2}{\Sigma}\right)
%	\right]
	%\partial_r g_{\phi\phi}
,~~~
\nonumber\\
&&
\Gamma^r_{\theta\phi}=
\Gamma^r_{\phi\theta}=\frac{1}{2}(	g^{tr}\partial_\theta g_{t\phi} +
		g^{rr}\partial_\theta g_{r\phi} +
		g^{r\phi}\partial_\theta g_{\phi\phi}
	) 
%\nonumber\\
%&&
=0%-%\frac{1}{2} 
	%\frac{a}{2\Sigma} %g^{tr}
	%frac{4m^2r^2(r^2+a^2)}{\Sigma^2}\sin 2\theta %\partial_\theta g_{t\phi} 
	%-
	%\frac{1}{2} 
	%\frac{\Delta}{\Sigma} %g^{rr}
	%a\sin 2\theta 
	%\left[
	%	1+\frac{2mr(r^2+a^2)}{\Sigma^2}
	%\right]
	%\partial_\theta g_{r\phi} 
	%+
	%\frac{1}{2} 
	%\frac{2mar}{\Sigma}\sin 2\theta  %g^{r\phi}
	%\left[
		%-\frac{4m^2r^2(r^2+a^2)}{\Sigma^2}
		%2mr
		%-\Delta
			%\left[
		%		-\frac{2mr(r^2+a^2)^2}{\Sigma^2}
			%\right]
		%+\Sigma %\sin 2\theta 
		%+a^2\sin^2\theta %\sin 2\theta 
			%\left[
		%		+\frac{2ma^2r\sin^2\theta}{\Sigma}
		%		\left(1+\frac{r^2+a^2}{\Sigma}\right)
			%\right]
	%\right]
	%\partial_\theta g_{\phi\phi}
,~~~
\nonumber\\
&&
\Gamma^\theta_{tt}=-\frac{1}{2}g^{\theta \theta}\partial_\theta g_{tt}
=%\frac{1}{2}
	%\frac{1}{\Sigma} %g^{\theta \theta}
	-\frac{ma^2r}{\Sigma^3}\sin 2\theta 
	%\partial_\theta g_{tt}
,~~~
\Gamma^\theta_{rr}=-\frac{1}{2}g^{\theta \theta}\partial_\theta g_{rr}
=%\frac{1}{2}
	%\frac{1}{\Sigma} %g^{\theta \theta}
	-\frac{ma^2r}{\Sigma^3}\sin 2\theta 
	%\partial_\theta g_{rr}
,~~~
%\nonumber\\
%&&
\Gamma^\theta_{\theta\theta}=\frac{1}{2} g^{\theta\theta} 
		\partial_\theta g_{\theta\theta}
=%\frac{1}{2} 
	-\frac{a^2}{2\Sigma}\sin 2\theta %g^{\theta\theta} 
	%2r %\partial_\theta g_{\theta\theta}
,~~~
\nonumber\\
&&
\Gamma^\theta_{\phi\phi}=
	-\frac{1}{2}g^{\theta \theta}\partial_\theta g_{\phi\phi}
=%-%\frac{1}{2}
	-\frac{\sin 2\theta}{2\Sigma} %g^{\theta \theta}
	%\sin^2\theta
	\left[
		\Delta + 2mr 
				\left(\frac{r^2+a^2}{\Sigma}\right)^2
	\right] %\partial_\theta g_{\phi\phi}
,~~~
%\nonumber\\
%&&
\Gamma^\theta_{tr}=
\Gamma^\theta_{rt}= -\frac{1}{2} g^{\theta \theta} \partial_\theta g_{tr}
= %\frac{1}{2} 
	%\frac{1}{\Sigma} %g^{\theta \theta} 
	-\frac{ma^2r}{\Sigma^3}\sin 2\theta 
	%\partial_\theta g_{tr}
,~~~
\nonumber\\
&&
\Gamma^\theta_{t\theta}=
\Gamma^\theta_{\theta t}=0 ,~~~
%\nonumber\\
%&&
\Gamma^\theta_{t\phi}=
\Gamma^\theta_{\phi t}=-\frac{1}{2} 
	g^{\theta \theta} \partial_\theta g_{t\phi}
=%\frac{1}{2} 
	%\frac{1}{\Sigma} %g^{\theta \theta} 
	\frac{mar(r^2+a^2)}{\Sigma^3}\sin 2\theta 		
	%\partial_\theta g_{t\phi}
,~~~
%\nonumber\\
%&&
\Gamma^\theta_{r\theta}=
\Gamma^\theta_{\theta r}=\frac{1}{2} 
	g^{\theta\theta} \partial_r g_{\theta\theta}
=%\frac{1}{2} 
	\frac{r}{\Sigma} %g^{\theta\theta} 
	%2r %\partial_r g_{\theta\theta}
,~~~
\nonumber\\
&&
\Gamma^\theta_{r\phi}=
\Gamma^\theta_{\phi r}= -\frac{1}{2} 
	g^{\theta \theta} \partial_\theta g_{r\phi}
= %\frac{1}{2} 
	%\frac{1}{\Sigma} %g^{\theta \theta} 
	\frac{a}{2\Sigma}\left[1+\frac{2mr(r^2+a^2)}{\Sigma^2}\right]\sin 2\theta
	%\partial_\theta g_{r\phi}
,~~~
\Gamma^\theta_{\theta\phi}=
\Gamma^\theta_{\phi\theta}=0 ,~~~
%%%%%%%%%%%%%%%%%%%%%%%%%%%%%%%%%%%%%%%%%%%%%%%%%%%%%%%%%%%%%%%%%%%%%%%%%%%
\nonumber\\
&&
\Gamma^\phi_{tt}=-\frac{1}{2}g^{r\phi}\partial_r g_{tt}
=-%\frac{1}{2}
	%\frac{a}{\Sigma} %g^{r\phi}
	\frac{ma}{\Sigma^2}\left(1-\frac{2r^2}{\Sigma}\right) %\partial_r g_{tt}
,~~~
\Gamma^\phi_{rr}=\frac{1}{2} g^{r\phi}\partial_r g_{rr} + 
		g^{\phi\phi}\partial_r g_{r\phi}
=-%\frac{1}{2} 
	%\frac{a}{\Sigma} %g^{r\phi}
	\frac{ma}{\Sigma^2}\left(1-\frac{2r^2}{\Sigma}\right) %\partial_r g_{rr} 
	%- 
	%\frac{1}{\Sigma} %g^{\phi\phi}
	%\frac{2ma}{\Sigma^2}\left(1-\frac{2r^2}{\Sigma}\right)
	%\partial_r g_{r\phi}
,~~~
\nonumber\\
&&
\Gamma^\phi_{\theta\theta}=
	-\frac{1}{2}g^{r\phi}\partial_r g_{\theta\theta}
=-%\frac{1}{2}
	\frac{ar}{\Sigma} %g^{r\phi}
	%2r %\partial_r g_{\theta\theta}
,~~~
\Gamma^\phi_{\phi\phi}=-\frac{1}{2}g^{r\phi}\partial_r g_{\phi\phi}
=-%\frac{1}{2}
	\frac{a}{\Sigma} %g^{r\phi}
	\sin^2\theta
	\left[
		r + \frac{ma^2\sin^2\theta}{\Sigma}
				\left(1-\frac{2r^2}{\Sigma}\right)
	\right]
	%\partial_r g_{\phi\phi}
,~~~
\nonumber\\
&&
\Gamma^\phi_{tr}=
\Gamma^\phi_{rt}=\frac{1}{2} g^{\phi\phi}\partial_r g_{t\phi}
=-%\frac{1}{2} 
	%\frac{1}{\Sigma\sin^2\theta} %g^{\phi\phi}
	\frac{ma}{\Sigma^2}\left(1-\frac{2r^2}{\Sigma}\right)
	%\partial_r g_{t\phi}
,~~~
%\nonumber\\
%&&
\Gamma^\phi_{t\theta}=
\Gamma^\phi_{\theta t}=\frac{1}{2} (	g^{r\phi}\partial_\theta g_{tr} + 
		g^{\phi\phi}\partial_\theta g_{t\phi}
	)
=-\frac{2mar}{\Sigma^2}\cot\theta
%\frac{1}{2} 
	%\frac{a}{\Sigma} %g^{r\phi}
	%\frac{ma}{\Sigma^2}\left(1-\frac{2r^2}{\Sigma}\right)
	%\partial_\theta g_{tr} 
	%- 
	%\frac{1}{2}
	%\frac{1}{\Sigma} %g^{\phi\phi}
	%\frac{ma}{\Sigma^2}\left(1-\frac{2r^2}{\Sigma}\right)
	%\partial_r g_{t\phi}
,~~~
\nonumber\\
&&
\Gamma^\phi_{t\phi}=
\Gamma^\phi_{\phi t}= -\frac{1}{2}
	g^{r\phi}\partial_r g_{t\phi}
= %\frac{1}{2}
	%\frac{a}{\Sigma} %g^{r\phi}
	\frac{ma^2\sin^2\theta}{\Sigma^2}\left(1-\frac{2r^2}{\Sigma}\right)
	%\partial_r g_{t\phi}
,~~~
%\nonumber\\
%&&
\Gamma^\phi_{r\theta}=
\Gamma^\phi_{\theta r}=\frac{1}{2} (	g^{r\phi}\partial_\theta g_{rr} + 
		g^{\phi\phi}\partial_\theta g_{r\phi}
	)
=-\frac{a}{\Sigma}\left(1+\frac{2mr}{\Sigma}\right)\cot\theta
%\frac{1}{2} 
	%\frac{a}{\Sigma} %g^{r\phi}
	%\frac{ma}{\Sigma^2}\left(1-\frac{2r^2}{\Sigma}\right)
	%\partial_\theta g_{rr} 
	%- 
	%\frac{1}{2}
	%\frac{1}{\Sigma} %g^{\phi\phi}
	%\frac{ma}{\Sigma^2}\left(1-\frac{2r^2}{\Sigma}\right)
	%\partial_r g_{r\phi}
,~~~
\nonumber\\
&&
\Gamma^\phi_{r\phi}=
\Gamma^\phi_{\phi r}=\frac{1}{2} g^{\phi\phi}\partial_r g_{\phi\phi}
=
\frac{1}{\Sigma}\left[
	r+\frac{ma^2\sin^2\theta}{\Sigma}
		\left(
			1-\frac{2r^2}{\Sigma}
		\right)
\right]
%-%\frac{1}{2} 
	%\frac{a}{\Sigma} %g^{r\phi}
	%\frac{ma^2\sin^2\theta}{\Sigma^2}\left(1-\frac{2r^2}{\Sigma}\right)
	%\partial_r g_{r\phi} 
	%+
	%\frac{1}{2}
	%\frac{r}{\Sigma} %g^{\phi\phi}
	%\sin^2\theta
	%\left[
	%	r + \frac{ma^2\sin^2\theta}{\Sigma}
	%			\left(1-\frac{2r^2}{\Sigma}\right)
	%\right]
	%\partial_r g_{\phi\phi}
,~~~
\nonumber\\
&&
\Gamma^\phi_{\theta\phi}=
\Gamma^\phi_{\phi\theta}
=\frac{1}{2} ( g^{r\phi}\partial_\theta g_{r\phi} 
				+ g^{\phi\phi}\partial_\theta g_{\phi\phi} )
=%\frac{1}{2} 
	%\frac{1}{\Sigma} %g^{\phi\phi}
	%\sin^2\theta
	\left[
		1 + \frac{2mr}{\Sigma}
				\left(\frac{r^2+a^2}{\Sigma}-1\right)
	\right]\cot\theta
	%\partial_r g_{\phi\phi}
.
\end{eqnarray}
%%%%%
%


\bsp

\label{lastpage}

\end{document}